\documentclass{emulateapj}
\usepackage{lscape, rotating}

\newcommand{\lapprox }{{\lower0.8ex\hbox{$\buildrel <\over\sim$}}}
\newcommand{\gapprox }{{\lower0.8ex\hbox{$\buildrel >\over\sim$}}}

\slugcomment{The Astrophysical Journal Supplement Series, 181:444-465, 2009 April}
\shorttitle{X-RAY-EMITTING STARS IDENTIFIED FROM THE RASS/SDSS}
\shortauthors{AG\"UEROS ET AL.}

\def\amin{\ifmmode^{\prime}\else$^{\prime}$\fi}
\def\asec{\ifmmode^{\prime\prime}\else$^{\prime\prime}$\fi}
\def\ROSAT{\it ROSAT}

\begin{document}

\title{X-ray Emitting Stars Identified From The $\ROSAT$ All-Sky Survey And The Sloan Digital Sky Survey\altaffilmark{1}}

\author{
Marcel A.\ Ag\"ueros\altaffilmark{2,3},
Scott F.\ Anderson\altaffilmark{4},
Kevin R.\ Covey\altaffilmark{5},
Suzanne L.\ Hawley\altaffilmark{4},
Bruce Margon\altaffilmark{6},
Emily R.\ Newsom\altaffilmark{3},
Bettina Posselt\altaffilmark{7},
Nicole M.\ Silvestri\altaffilmark{4},
Paula Szkody\altaffilmark{4},
Wolfgang Voges\altaffilmark{8}
}

\altaffiltext{1}{Includes observations obtained with the Apache Point Observatory $3.5$-m telescope, which is owned and operated by the Astrophysical Research Consortium.}
\altaffiltext{2}{NSF Astronomy and Astrophysics Postdoctoral Fellow; marcel@astro.columbia.edu}
\altaffiltext{3}{Columbia Astrophysics Laboratory, Columbia University, 550 West 120th Street, New York, NY 10027}
\altaffiltext{4}{Department of Astronomy, University of Washington, Box 351580, Seattle, WA 98195}
\altaffiltext{5}{Spitzer Fellow, Harvard-Smithsonian Center for Astrophysics, 60 Garden Street, Cambridge, MA 02138}
\altaffiltext{6}{Department of Astronomy and Astrophysics, University of California, 1156 High Street, Santa Cruz, CA 95064}
\altaffiltext{7}{Leopoldina Fellow, Harvard-Smithsonian Center for Astrophysics, 60 Garden Street, Cambridge, MA 02138}
\altaffiltext{8}{Max-Planck-Institut f\"ur extraterrestrische Physik, Geissenbachstrasse 1, D-85741 Garching, Germany} 

\begin{abstract}
The $\ROSAT$ All-Sky Survey (RASS) was the first imaging X-ray survey of the entire sky. Combining the RASS Bright and Faint Source Catalogs yields an average of about three X-ray sources per square degree. However, while X-ray source counterparts are known to range from distant quasars to nearby M dwarfs, the RASS data alone are often insufficient to determine the nature of an X-ray source. As a result, large-scale follow-up programs are required to construct samples of known X-ray emitters. We use optical data produced by the Sloan Digital Sky Survey (SDSS) to identify $709$ stellar X-ray emitters cataloged in the RASS and falling within the SDSS Data Release~1 footprint. Most of these are bright stars with coronal X-ray emission unsuitable for SDSS spectroscopy, which is designed for fainter objects ($g > 15$ mag). Instead, we use SDSS photometry, correlations with the Two Micron All Sky Survey and other catalogs, and spectroscopy from the Apache Point Observatory $3.5$-m telescope to identify these stellar X-ray counterparts. Our sample of $707$ X-ray emitting F, G, K, and M stars is one of the largest X-ray selected samples of such stars. We derive distances to these stars using photometric parallax relations appropriate for dwarfs on the main sequence, and use these distances to calculate $L_X$. We also identify a previously unknown cataclysmic variable (CV) as a RASS counterpart. Separately, we use correlations of the RASS and the SDSS spectroscopic catalogs of CVs and white dwarfs (WDs) to study the properties of these rarer X-ray emitting stars. We examine the relationship between $(f_X/f_g)$ and the equivalent width of the H$\beta$ emission line for $46$ X-ray emitting CVs and discuss tentative classifications for a subset based on these quantities. We identify $17$ new X-ray emitting DA (hydrogen) WDs, of which three are newly identified WDs. We report on follow-up observations of three candidate cool X-ray emitting WDs (one DA and two DB (helium) WDs); we have not confirmed X-ray emission from these WDs.
\end{abstract}


\section{Introduction}
X-ray data obtained since the 1970s have shown definitively that stellar X-ray emitters are present among almost all stellar classes \citep[e.g.,][]{harnden79, cass83, stocke83, schmitt95, motch98, zickgraf03, rogel06}, with the emission mechanism being different for early and late-type stars; only evolved K and M stars and perhaps A stars do not appear to be X-ray emitters \citep[e.g.,][]{ayres81, antiochos86, haisch91, daniel02}. However, within a stellar class--particularly among late-type stars--the strength of the X-ray emission varies greatly. A G5 star may be much brighter in the X-ray regime than its apparent twin, suggesting that the physics of X-ray emission is intimately linked to detailed stellar properties. X-ray spectroscopy with the {\it Chandra X-ray Observatory} and {\it XMM-Newton X-ray Observatory}, occasionally in conjunction with observations at other wavelengths, is providing unprecedented insight into these connections \citep[e.g.,][]{osten05,smith05,hussain07}. Before such studies can be conducted, however, catalogs of confirmed X-ray emitting stars from all areas of the Hertzsprung-Russell diagram need to be compiled and broadly characterized.

We have identified a sample of stellar X-ray emitters cataloged in the $\ROSAT$ All-Sky Survey \citep[RASS;][]{voges99} and falling within the footprint of the Sloan Digital Sky Survey \citep[SDSS;][]{york00} Data Release~1 \citep[DR1;][]{dr1}. We produced a two-part catalog. One part includes $231$ stars that are previously cataloged, e.g., in SIMBAD or in the NASA/IPAC Extragalactic Database (NED)\footnote{This research has made use of the SIMBAD database, which is operated at CDS, Strasbourg, France, and of the NASA/IPAC Extragalactic Database and Infrared Science Archive, which are operated by the Jet Propulsion Laboratory, California Institute of Technology, under contract with the National Aeronautics and Space Administration.}, but not identified as X-ray source counterparts. Another $478$ entries are new stellar X-ray source identifications made on the basis of spectroscopic observations with the Astrophysical Research Consortium $3.5$-m telescope at the Apache Point Observatory (APO). The vast majority of these stars are F, G, K, and M stars ($707$ stars), although we identify a new cataclysmic variable (CV) in our APO data; the SIMBAD sample also includes a newly identified X-ray emitting white dwarf (WD). To expand the sample of X-ray emitting CVs and WDs, we correlated the SDSS spectroscopic catalogs of these objects \citep[][]{paula1, paula2, paula3, paula4, paula5, paula6, eisenstein06} with the RASS to identify the X-ray emitters among these more exotic stars. In a previous paper we described our efforts to use the RASS and SDSS to identify new isolated neutron stars, which are possibly the rarest stellar X-ray emitters \citep{me}.

In Section~\ref{hist_IDs}, we review other efforts to construct large samples of X-ray emitting stars; we also provide a brief description of SDSS, as well as a summary of the challenges of using SDSS to identify RASS stellar X-ray emitters. In Section~\ref{selection}, we outline the process by which we constructed our list of candidate DR1 stellar counterparts to RASS sources. In Section~\ref{sample}, we describe how we confirmed our stellar identifications, thereby obtaining our sample of stellar X-ray sources. We distinguish between previously cataloged stars we identify as X-ray sources and stars for which we obtained spectra with the APO $3.5$-m telescope. In Section~\ref{master_cat}, we begin by calculating distances and X-ray luminosities for $662$ stars in our catalog, compare our identifications to those of \citet{flesch04} and of \citet{parejko08} for sources that also appear in their catalogs, and describe the data in our main catalog, which includes the X-ray, optical, and infrared information for our $709$ stellar X-ray sources. We also estimate the overall reliability of our identifications; we find that $\lapprox\ 10\%$ are likely to be false associations. In Section~\ref{wds_cvs}, we present the results of a parallel effort to use correlations of the SDSS spectroscopic catalogs of CVs and of WDs with the RASS to construct catalogs of these rarer stellar X-ray emitters. We also report on follow-up {\it Chandra} and {\it XMM} observations of several candidate cool X-ray emitting WDs. We conclude in Section~\ref{concl}.

\section{Identifying Stellar X-ray Sources...}\label{hist_IDs}

\subsection{using optical photometry and spectroscopy}
The {\it Einstein Observatory} provided the first large samples of X-ray sources. The Medium Sensitivity Survey \citep[MSS;][]{gioia84} and the Extended Medium-Sensitivity Survey \citep[EMSS;][]{gioia90} collected serendipitously detected sources with fluxes between $\sim5 \times 10^{-14}$ and $3 \times 10^{-12}$~ergs~cm$^{-2}$~s$^{-1}$ in the $0.3 - 3.5$~keV energy band. Because survey X-ray data are often insufficient to determine unambiguously the nature of a source, extensive optical follow-up was required to identify the $112$ MSS and the $835$ EMSS sources. \citet{stocke83} used the MSS to confirm that a given class of X-ray emitters has a fairly narrow range of possible X-ray-to-optical flux ratios (e.g., for active galactic nuclei (AGN), $-1.0 \leq $~log $(f_X/f_V) \leq +1.2$), and that the overlap between broad classes (e.g., Galactic and extragalactic sources) is fairly small. They found that $\sim 25\%$ of MSS sources were coronally emitting stars, primarily late-type dwarfs; they also found one cataclysmic variable (CV). \citet{stocke91} used the MSS results to identify plausible counterparts to EMSS sources based on their $(f_X/f_{opt})$. Still, confirming that these counterparts were X-ray sources was very time-intensive: \citet{stocke91} obtained two to five spectra in each X-ray error circle. Six to eight nights a year over seven years were needed at the Multiple Mirror Telescope Observatory, Mt.\ Hopkins, AZ, to identify $665$ sources. This work confirmed that $\sim25\%$ of {\it Einstein} detected-sources are Galactic stars.

Similar efforts have been undertaken to identify some of the $\sim125,000$ sources detected at soft X-ray energies ($0.1 - 2.4$ keV) by $\ROSAT$ as part of its All-Sky Survey. The RASS Bright Source Catalog \citep[BSC;][]{voges99} includes $18,811$ sources with count rates $\ge 0.05$~cts~s$^{-1}$ and $\geq 15$ counts; the Faint Source Catalog \citep[FSC;][]{fsc} lists another $105,924$ with $< 0.05$ cts~s$^{-1}$. A relatively small fraction of RASS sources can be identified from correlations with existing catalogs. \citet{bade98} found matches to $35\%$ of a sample of $80,000$ RASS sources in SIMBAD and NED. To identify other BSC sources, \citet{bade98} used objective prism spectra obtained during the Hamburg Quasar Survey \citep[HQS;][]{hagen95} and found candidate counterparts for $81.2\%$ of the $3847$ sources within the HQS footprint. Of these, $155$ ($4\%$) were M stars, $136$ ($3.5\%$) K stars, $4$ ($0.1\%$) F or G stars; a further $956$ ($24.9\%$) were saturated bright stars ($B \leq 14$ mag), for which no spectral class is available. There were also $31$ WDs ($0.8\%$), and $16$ CVs ($0.4\%$). There are uncertainties associated with these identifications; the spectra are fairly low-resolution, for example (R $\approx 100$ at H$\gamma$). But the RASS/HQS program suggests that one-third of $\ROSAT$-detected sources are Galactic stars.

\citet{zickgraf03} applied this method to identify all $5341$ RASS BSC sources with $\delta \geq 0^{\circ}$ and Galactic latitudes $\left|b\right| \geq 30^{\circ}$; the total area covered was $10,313$ deg$^2$. Candidate counterparts were found for $82.2\%$ of the RASS sources; the ``missing'' counterparts are faint optical objects whose poor-quality prism spectra preclude classification. The fraction of the $5341$ BSC sources due to coronal emission from stars increases from $33\%$ to $\sim38\%$ when one includes the likely stellar sources among the missing counterparts \citep{zickgraf03}. The difference between this fraction and that of stellar sources in the EMSS may be due to the different energy sensitivities of $\ROSAT$ and {\it Einstein} \citep{zickgraf03}. The final RASS/HQS sample includes $197$ M stars, $141$ K stars, $45$ F/G stars, and $1219$ saturated stars without spectral classifications, along with $45$ WDs and $26$ CVs.

\subsection{using catalog matching}
Parallel efforts have been made to produce large-scale catalogs of RASS counterparts without spectroscopy. \citet{voges99} correlated the $18,811$ BSC sources with a number of existing catalogs--for example, of bright stars \citep[Tycho;][]{tycho} or of radio and ultraviolet (UV) sources \citep[FIRST, NVSS, and EUVE;][]{first,nvss,euve}. They found at least one cataloged counterpart within $90''$ for $90\%$ of BSC sources. $7117$ X-ray sources had a single cataloged counterpart, which \citet{voges99} used to establish the overall statistical properties of the catalog: $58\%$ have Galactic counterparts. This matching was based primarily on positional coincidence and relied on the relatively low surface densities of BSC sources and of the various candidate counterparts \citep{voges99}.

\citet{rutledge00} matched the BSC to the USNO A$-2$ catalog \citep{usnoa2}. Factoring in both the proximity and the USNO $B$ of objects found within $75''$ of the RASS sources, and estimating the contamination rate by comparisons to background fields, they sifted through $> 320,000$ possible optical counterparts to identify those with the highest probability of being X-ray sources. The resulting catalog includes $11,301$ BSC sources for which \citet{rutledge00} identified a USNO object with a $\geq 50\%$ probability of being the counterpart; $2705$ had a probability $\geq 98\%$.

In an even more ambitious effort, \citet{flesch04} matched X-ray data from all of the major $\ROSAT$ catalogs to multiple radio and optical surveys, thereby creating an all-sky catalog of radio/optical counterparts to some $500,000$ X-ray sources. Nearly $120,000$ of these sources were previously identified, while an additional $86,000$ are estimated to be $40\%$ to $99\%$ likely to be previously unknown quasars (QSOs). The likelihood of a given association was calculated first by examining the proximity of a proposed counterpart, its $B-R$ color, and its optical classification, and then by comparing the density of all candidate counterparts with these properties to their all-sky density \citep{flesch04}. 

\vspace{-.25cm}
\begin{figure}[th]
\epsscale{1.1}
\centerline{\plotone{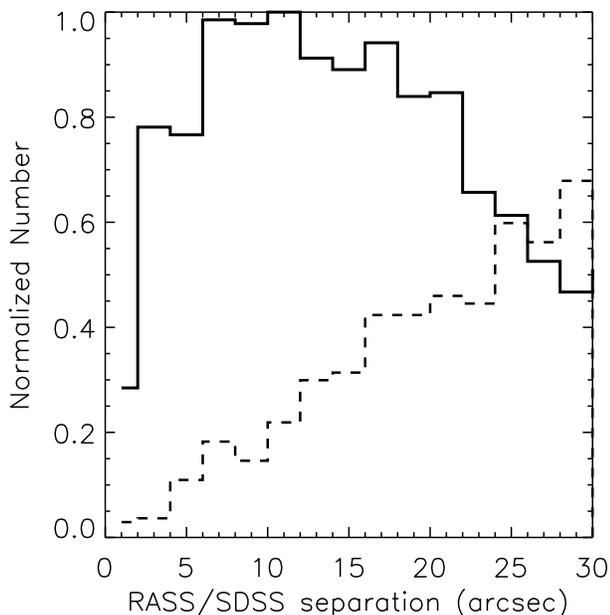}}
\caption{{\it Solid line:} distribution of separations between RASS sources and nearby bright/saturated SDSS objects with 2MASS counterparts. {\it Dashed line:} distribution of separations returned by matching shifted X-ray sources to SDSS/2MASS objects.}\label{matching}
\end{figure}

\vspace{-.25cm}
These catalogs are very useful for developing a statistical sample of a given category of X-ray emitters, typically QSOs \citep[e.g,][]{bauer00, flesch04}. (There is usually almost no discussion of the stellar X-ray sources.) However, because they rely largely on the properties of known counterparts to identify new counterparts, the catalogs cannot tell us more about the overall properties of X-ray sources than the smaller catalogs on which they are based \citep[][]{flesch04}. In that sense, these catalog-based approaches cannot fully replace the combined photometric and spectroscopic method developed to identify {\it Einstein} sources. 

Furthermore, detailed analysis of selected subsets of these catalogs can reveal problems with the claimed identifications. For example, intrigued by apparent anomalies in the spatial distribution of QSOs in the \citet{flesch04} catalog, \citet{lopezc08} investigated a subsample of $41$ QSO candidates to test the reliability of their assigned QSO likelihoods. They found that only $12$ were actual QSOs, and that the catalog likelihoods were particularly unreliable for bright objects ($B \leq 15.1$ mag), where none of the $13$ \citet{flesch04} candidates was in fact a QSO.

\subsection{using SDSS}\label{surveys}
Previous to this work, the \citet{zickgraf03} catalog was the single largest sample of RASS stellar sources for which counterparts have been spectroscopically confirmed. In fact, only $\sim10\%$ of RASS BSC/FSC sources are identified and fully characterized \citep{mcglynn04,mick06}, with the majority of the identifications being of BSC sources. Large-scale programs to find the counterparts to RASS sources are still very much needed.

SDSS provides uniform optical photometric and spectroscopic data with which to correlate the RASS catalogs. The survey produces {\it u, g, r, i, z} images to a depth of $r\sim22.5$ mag \citep{fukugita, gunn, hogg01, smith02, gunn06}, with a photometric accuracy of $\sim 0.02$ mag \citep{zeljko04}; \citet{tucker06} describe the photometric pipeline. The imaging survey covers $8420$~deg$^2$ and includes photometric data for over $10^8$ stars and a similar number of galaxies. SDSS is also a spectroscopic survey, and has obtained spectra for $\sim8\times10^5$ galaxies, $\sim10^5$ QSOs, and $\sim2\times10^5$ stars. 

SDSS is an ideal tool for systematically identifying large numbers of $\ROSAT$ sources \citep[e.g., X-ray emitting AGN or galaxies;][]{anderson03,anderson06,parejko08}. However, the vast majority of stellar X-ray emitters cataloged in the RASS are unlikely to be discovered from routine SDSS spectroscopy. Relative to their optical output, the most X-ray luminous normal stars (i.e., coronal emitters) have log $(f_X/f_{opt})$~$\sim-1$ \citep{macca88} and so almost all are optically brighter than the $g\approx15$ mag SDSS spectroscopic limit. 

\begin{figure*}[th]
\centerline{\includegraphics[width=1.95\columnwidth]{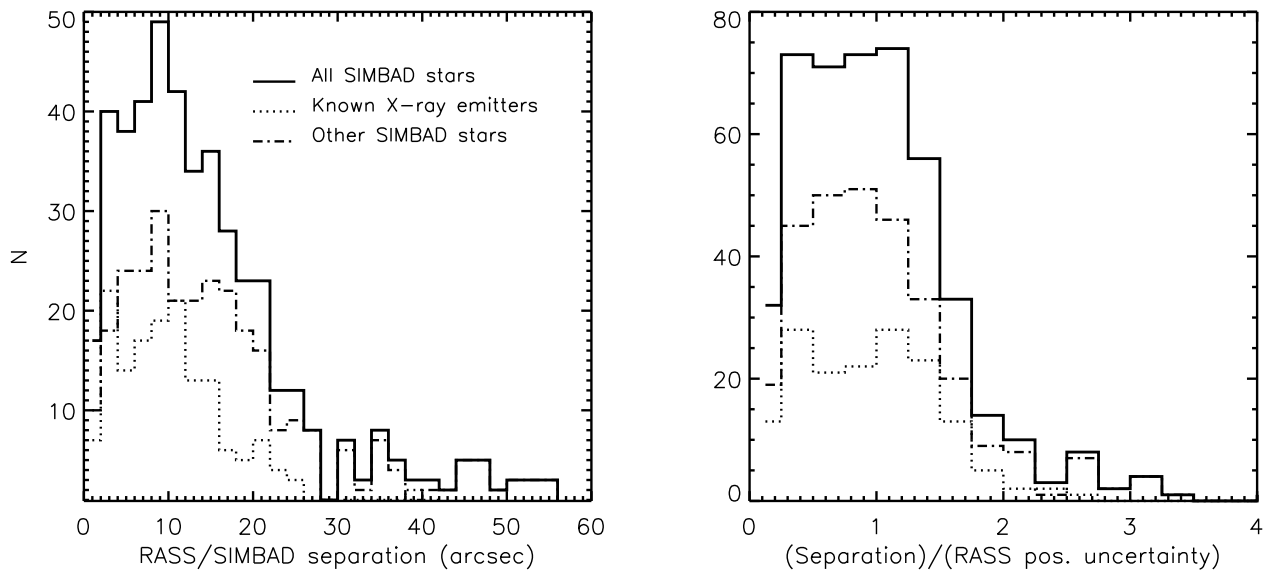}}
\caption{{\it Left panel:} distribution of positional offsets between the RASS sources and the SIMBAD stars. {\it Right panel:} the same distribution, expressed in terms of the cataloged X-ray positional uncertainty.}\label{separation}
\end{figure*}

We have therefore combined the SDSS photometric data with infrared data from the Two Micron All Sky Survey \citep[2MASS;][]{2mass} to select a sample of bright stars near BSC/FSC positions. When available, we used existing catalog data to classify the stars. In many cases, however, no spectral class is available, and we therefore used the APO $3.5$-m telescope to obtain spectra for objects in about $900$ of these RASS error circles (the total number of spectra collected is close to $1200$). We then used a spectral-template fitting code to classify the likely counterparts by spectral type. We also estimated $(f_X/f_{opt})$ ratios for all the stars based on their cataloged magnitudes and compared the ratios to those of known X-ray emitters of the same spectral type, thereby verifying that these stars are indeed plausible RASS source counterparts. Each of these steps is described in detail in the following sections.

\section{Selecting X-ray Emitting Stars from the RASS and SDSS DR1}\label{selection}
We matched the $\sim125,000$ RASS X-ray source positions to those of the SDSS Data Release 1 \citep[DR1;][]{dr1} stripes; each SDSS stripe is a $2.5$-deg-wide rectangle of sky. We found matches for $7464$ RASS sources, or just under $6\%$ of the catalogs' entries. Since the DR1 area is $2099$~deg$^2$ \citep{dr1}, or roughly $5\%$ of the entire sky, this number is consistent with a naive surface density expectation\footnote{The quoted DR1 area is for the imaging survey; the DR1 spectroscopic area is $1360$~deg$^2$.}.

We queried the SDSS photometric database using the SDSS Query Analyzer (sdssQA) to identify bright stars near these RASS sources; see \citet[][]{thesis} for details. This query returned coordinates, PSF magnitudes\footnote{PSF fitting provides better estimates of isolated star magnitudes; see \citet{stoughton02}.}, and automatically assigned morphology type (star or galaxy) for objects with $g$ or $r$ $<16$ mag, subject to a small number of tests of the objects' photometric flags.

The output from sdssQA was matched to the RASS sources in the DR1 footprint. This matching produced a list of $\lapprox\ 5000$ RASS sources with a bright/saturated SDSS object within $1'$, roughly two-thirds of the total number of RASS/DR1 sources. However, visual inspection of the SDSS images for these sources revealed that a significant fraction did not have a nearby bright object\footnote{We used the SDSS image list tool to examine our error circles \citep{sdss_images}.}. While a more conservative set of photometric cuts might eliminate many contaminants, it might also remove the real but saturated objects in which we are interested. We therefore used the Two Micron All Sky Survey \citep[2MASS;][]{2mass} to confirm the existence of bright SDSS objects in the RASS error circles. While SDSS is a much deeper survey than 2MASS, whose $J$, $H$, and $K_s$ limits are $\sim14$ mag, the two surveys are well matched at the bright end ($r \leq 16$), and our target sample of bright SDSS objects should be detected by 2MASS.

Accordingly, the 2MASS catalog was queried for matches to the $5000$ SDSS objects. The typical positional uncertainties for point sources in both surveys are on the order of a fraction of an arcsecond \citep[e.g.,][]{finlator,pier}; the median separation between matches was $0.22''$. This matching left us with a list of $2520$ bright/saturated SDSS objects with a nearby 2MASS counterpart within $1'$ of a RASS source. In Figure~\ref{matching} we show the normalized histogram of the resulting RASS/SDSS separations $\leq 30''$.

To estimate the likelihood of false matches, we shifted the positions of RASS sources in the DR1 footprint by $2$\amin\ in declination and searched for nearby SDSS objects with 2MASS counterparts in the same manner as described above. Figure~\ref{matching} shows the (dashed) normalized histogram for this control sample. The number of false matches rises with separation, as expected. The normalization is the same as for the previous matches, so that the dashed histogram gives an upper limit to the contamination of our sample by chance superpositions of independent RASS sources and optical/IR objects. At $25$\asec, the cumulative contamination is $\sim35\%$. Absent other information, e.g., the brightness or nature of the proposed counterpart, positional proximity alone is generally insufficient to make secure RASS source identifications from SDSS/2MASS data. We return to the question of contamination by false matches in \S~\ref{false}, after we have assembled our final catalog. 

In general, the SDSS spectroscopic database does not provide the needed additional information to identify stellar RASS counterparts. As objects brighter than $\sim 15$ mag are rarely targeted, the RASS counterparts most likely to be targeted for spectroscopy are QSOs and other AGN, since they tend to be optically fainter than X-ray emitting stars \citep[cf.][]{anderson03,anderson06}. In 2003 Sep, $\sim680,000$ SDSS spectra were available, a sample we defined as making up the DR1 spectroscopic survey, yet only $\sim70$ of our matched RASS/SDSS/2MASS objects have an associated SDSS spectrum. Clearly, the majority of our RASS DR1 sources cannot therefore be directly identified from SDSS spectroscopy. 

\section{The RASS/SDSS DR1 Catalog of Stellar X-ray Sources}\label{sample}
To identify new stellar X-ray sources, we begin by searching SIMBAD to find stars with known spectral types recovered in our RASS/SDSS correlation. Among these are a subset of previously identified X-ray sources, to which we add the (optically bright) cataloged stars with small offsets relative to the RASS positions to construct flux ratio distributions. Using spectral types obtained from our APO observations, we then find those previously uncatalogued stars whose log~($f_X/f_J$) falls within the $2\sigma$ range obtained from our RASS/SIMBAD sample. Our final catalog includes these stars along with those cataloged in SIMBAD but not previously identified as RASS sources. The steps in constructing this catalog are described in detail below.

\subsection{RASS/SIMBAD stars}
We searched SIMBAD for cataloged candidate stellar counterparts to our $2520$ matched RASS sources and found $457$ stars {\it with an associated spectral type}. Examining the literature, we determined that $159$ of these stars were previously associated with $\ROSAT$ X-ray sources \citep[e.g., by][]{dempsey93,appenzeller98,fleming98,hunsch1998a,hunsch1998b,metanomski98,hunsch99,schwope,zickgraf03} and more rarely with sources detected by {\it Einstein} \citep[e.g., by][]{topka82,fleming89,fleming89b,stocke91,drake92} or by an even earlier X-ray mission \citep[e.g., the RS CVn systems DM UMa, detected by the {\it High Energy Astrophysics Observatory-1} and IN Vir, by {\it EXOSAT};][]{walter80, giommi91}. In four additional cases the stars were listed as the RASS source counterparts, but the origin of the association was unclear\footnote{These are FBS 0124$-$098, NLTT 6782, FBS 0249$-$084, and TYC 255$-$263$-1$.}. Conversely, $294$ of these SIMBAD stars had not been associated with an X-ray source.

Unsurprisingly, these stars tend to be optically very bright. $427$ have a cataloged $V$; for these, the median is $8.95$ mag. The stars are generally associated with brighter X-ray sources: the sample median count rate is $0.04$ counts~s$^{-1}$. By contrast, the median for the $2520$ RASS sources is $0.03$ counts~s$^{-1}$. The median separation between RASS sources and SIMBAD stars is $12.0$\asec; this is equivalent to $0.9\sigma$, where $\sigma$ is the separation between the matched positions divided by the RASS positional uncertainty (see Figure~\ref{separation}). The offset distribution of the known X-ray emitters in the sample suggests that the number of X-ray source counterparts drops quickly for separations $\gapprox\ 30''$ or $\gapprox\ 2\sigma$; only eight stars are at offsets $> 30$\asec; six have separations $> 2 \sigma$. 

The optical brightness and small positional offsets of the previously unpublished RASS/SIMBAD matches strongly suggest that they are also the X-ray sources. We replicate the matching to the RASS BSC done by \citet[][]{voges99} for stars in the Tycho catalog; the matched stars are comparable in brightness to our SIMBAD stars (their median $V \sim 10$ mag). For offsets $< 30$\asec\ the fraction of chance Tycho/RASS coincidences is $\sim 2\%$. This implies that $\sim 248$ of the $253$ stars with offsets $< 30''$ not yet associated with an X-ray source are likely RASS source counterparts. 

\begin{figure}[h]
\epsscale{1.20}
\centerline{\plotone{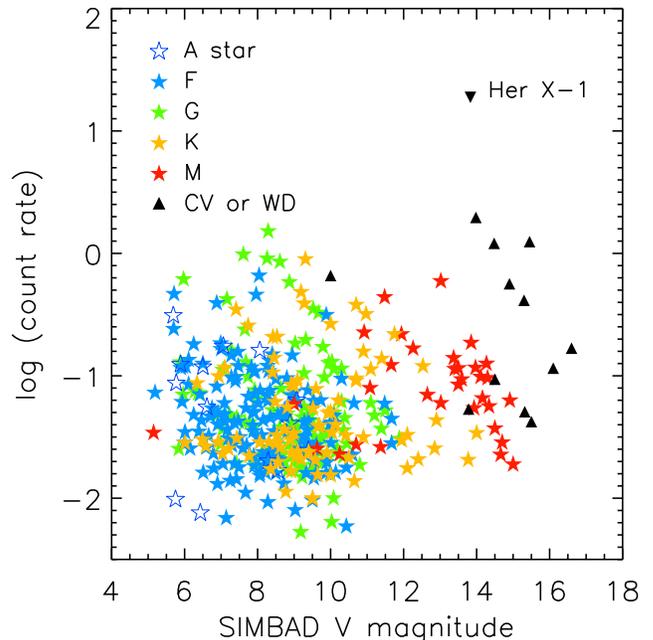}}
\caption{X-ray count rate vs.\ $V$ for $139$ known X-ray emitting SIMBAD stars and $249$ SIMBAD stars offset by $< 30$\asec\ from a RASS source. $B$ was used for the two CVs and two WDs with no cataloged $V$.}\label{known_counts}
\end{figure}

Figure~\ref{known_counts} displays the RASS count rate vs.\ SIMBAD $V$ for $139$ known X-ray emitters and for $249$ other SIMBAD stars with $< 30$\asec\ offsets\footnote{$20$ known X-ray emitting stars lack a cataloged $V$, as do four previously unidentified candidate RASS counterparts.}. These are primarily F, G, K, and M stars, although there are a few candidate A star counterparts (see the Appendix). There are also a handful of WDs (including a WD/M-star binary) and of CVs, which stand out because of their relative optical faintness but X-ray brightness. The brightest source is the low-mass X-ray binary Hercules X-1, among the first extrasolar X-ray sources to be detected \citep{schreier72, tananbaum72}.
 
\subsubsection{X-ray-to-optical flux ratio calculations} \label{ratio_calc}
Numerous studies have shown that there is a typical range of X-ray-to-optical flux ratios for each stellar type \citep[e.g.,][]{stocke91, krautter99, zickgraf03}. Accordingly, we calculate log $(f_X / f_V)$ for our stars. We exclude the $23$ stars listed in SIMBAD as belonging to luminosity classes II, III, or IV (see the Appendix). We also remove all stars for which the RASS match has an X-ray flag set to $1$, which indicates that the X-ray data are not reliable \citep[see][]{voges99}.

\begin{figure*}[t]
\centerline{\includegraphics[angle=90,width=1.7\columnwidth]{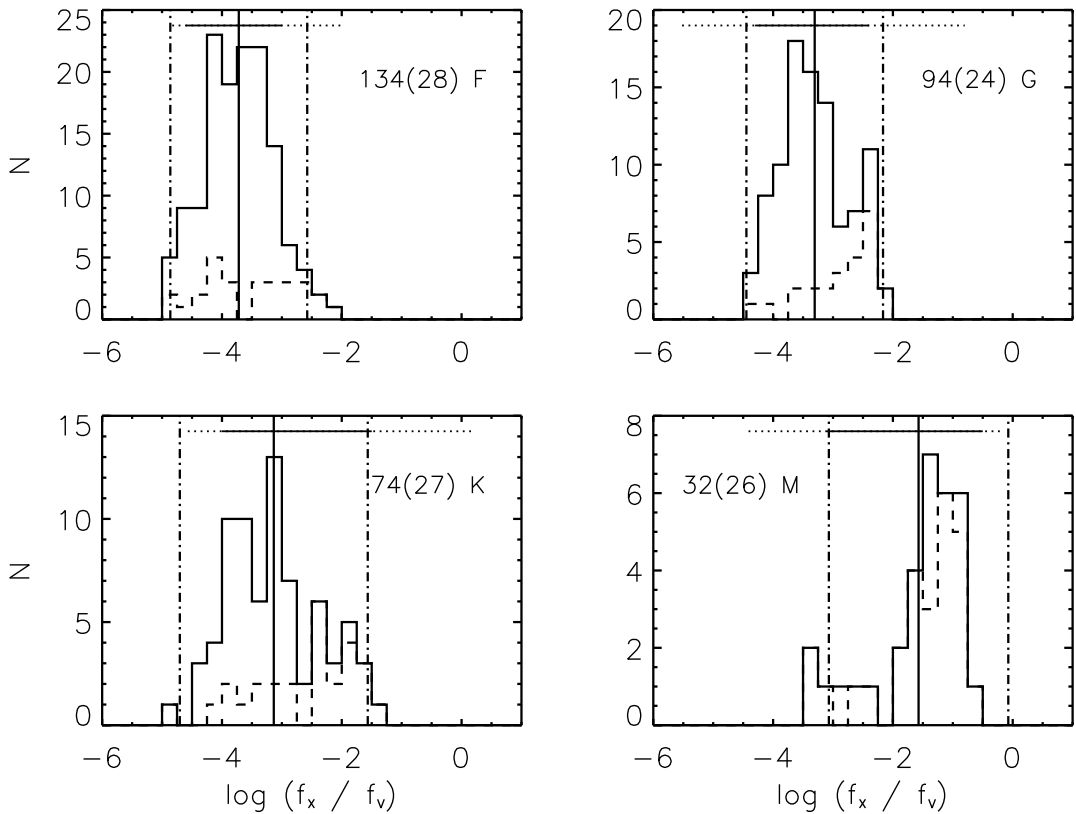}}
\caption{log $(f_X / f_V)$ distributions for the RASS/SIMBAD F, G, K, and M stars. The dashed histograms are the distributions for the subsets of previously known X-ray emitters. The solid vertical lines correspond to the mean $(f_X / f_V)$ and the dot-dashed vertical lines bracket the $2 \sigma$ $(f_X / f_V)$ range for each stellar type. The number of RASS/SIMBAD stars used to calculate these $(f_X / f_V)$ ranges is indicated in each panel; the number in parentheses is the subset of previously known X-ray emitters. The horizontal lines are $(f_X / f_V)$ ranges from the literature: the solid lines are from the \citet{macca88} {\it Einstein} survey, while the dotted extensions are from \citet{krautter99}.}\label{known_ratios}
\end{figure*}

We do not remove the known binary systems from this sample when calculating the flux ratios. There are $64$ such systems within $30$\asec\ of a RASS source and with RASS flags set to $0$; one more is a known X-ray emitter with an offset $> 30''$. $40$ are listed in SIMBAD as double or multiple star systems; one is in a star cluster, and one was identified as a binary through high-resolution spectroscopy by \citet{metanomski98}. Seven of these systems have been previously cataloged as X-ray sources. The sample also includes two Algol-type eclipsing and five spectroscopic binaries, four $\beta$ Lyr, six RS CVn, and six W UMa systems (including W UMa itself). While these systems may seem unrepresentative, many are known X-ray emitters (e.g., all the W UMa and RS CVn systems, but also the Algol-type binaries), which argues for their inclusion in flux ratio calculations\footnote{In practice, including the binaries broadens the flux ratio distributions slightly by adding objects at higher $(f_X / f_V)$.}.

As part of the spectroscopic campaign described in \S~\ref{spec_sample}, we obtained spectra for $40$ of the SIMBAD stars with listed spectral types. Only $20$ meet the criteria for inclusion in these calculations, and in only three cases would using our APO-based spectral classifications affect the $(f_X / f_V)$ distributions. In two cases, the difference in spectral type is consistent with our spectral typing uncertainty (see \S~\ref{spec_sample}): V* V842 Her, for which SIMBAD gives an F9V spectral type, and which we identify as a G2 star, and StKM $1-1262$, which \citet{stephenson86} classifies as a K5 star, and which we find is an M0. In both cases, we use the SIMBAD type. The third star is RX J$1057.1-0101$, classified as a K4V star by \citet{zickgraf05}. We type the RASS counterpart as a G5 star; however, examination of this error circle reveals that our APO target is not RX J$1057.1-0101$ but a nearby star. See the Appendix for further discussion of these stars.

We then have $105$ known X-ray emitting F, G, K, and M stars with a cataloged $V$ and with RASS flags $=0$, and an additional $229$ SIMBAD stars not previously associated with an X-ray source that meet the same criteria and are offset by $ < 30''$ from a RASS source. Following \citet{macca88}, we calculate log $(f_X /f_V) =$ log $(f_X) + 0.4 V + 5.37$ and assume that on average $1$ count~s$^{-1}$ translates to $f_X = 10^{-11}$~erg~cm$^{-2}$~s$^{-1}$ in the $0.1 - 2.4$ keV energy range \citep[e.g.,][]{motch98, voges99}. 

We first create separate distributions for the $105$ known X-ray emitters and for the other $229$ SIMBAD stars. Kolmogorov-Smirnov tests of the resulting flux ratio distributions find that for F, G, and K stars the log $(f_X / f_V)$ distributions are drawn from different underlying distributions ($P\approx0$), while for the M stars the distributions are weakly related ($P=0.6$; however, there are only six previously unidentified candidate M star counterparts). This is consistent with the fact that the known X-ray emitters are generally somewhat brighter in the optical and significantly brighter in the X ray than the SIMBAD stars not previously associated with a RASS source. Given the high probability that based on positional coincidence the $229$ previously unassociated stars considered here are RASS sources, we group the known and previously unknown X-ray emitters together, and calculate log $(f_X / f_V)$ for all of the stars of a given spectral type. The resulting distributions are presented in Figure~\ref{known_ratios}, as are empirical $2 \sigma$ ranges estimated from the mean log $(f_X / f_V)$ for each spectral type. 

\begin{deluxetable*}{lcclccccc}
\setlength{\tabcolsep}{0.5in}
\tabletypesize{\scriptsize}
\tablecaption{RASS/SIMBAD stars\label{known_table}}
\tablehead{
                         & \colhead{Initial}   & \colhead{New RASS} & \multicolumn{3}{c}{log $(f_X/f_V)$} & \multicolumn{3}{c}{log $(f_X / f_J)$}\\ 
\cline{4-9}
\multicolumn{1}{c}{Type} & \colhead{Number}    & \colhead{Sources}  & \colhead{Mean} & \colhead{$2\sigma$ Min} & \colhead{$2\sigma$ Max} & \colhead{Mean} & \colhead{$2\sigma$ Min} &\colhead{$2\sigma$ Max} 
}
\startdata
F   & $169\ (38)$ & $104$ & $-3.72$ & $-4.86$ & $-2.58$ & $-3.14$ & $-4.18$ & $-2.09$ \\
G   & $108\ (29)$ & $70$  & $-3.31$ & $-4.45$ & $-2.17$ & $-2.82$ & $-4.15$ & $-1.49$ \\
K   & $104\ (42)$ & $47$  & $-3.13$ & $-4.70$ & $-1.57$ & $-2.77$ & $-4.27$ & $-1.27$ \\
M   & $45\ (31)$  & $9$\tablenotemark{a}   & $-1.57$ & $-3.07$ & $-0.07$ & $-2.22$ & $-3.30$ & $-1.15$\\
\tableline
    &             &       & Mean    & Min     & Max\\
\tableline
A   & $18\ (7)$   & $0$   & $-3.86$ & $-5.34$ & $-2.28$ & \nodata & \nodata & \nodata\\
WD\tablenotemark{b}   & $8\ (7) $   & $1$   & $-0.04$ & $-0.86$ & $+0.65$ & \nodata & \nodata & \nodata\\
CV  & $4\ (4) $   & $0$   & $-0.73$\tablenotemark{c} & $-2.04$\tablenotemark{c} & $-0.12$\tablenotemark{c} &  \nodata & \nodata & \nodata\\
XRB & $1\ (1) $   & $0$   & $+1.18$ & \nodata & \nodata & \nodata & \nodata & \nodata
\enddata
\tablecomments{The number in parentheses in the second column is the number of known X-ray emitters. Empirical $2\sigma$ log $(f_X / f_V)$ ranges and means are given for reliable samples of F, G, K, and M stars. The full flux ratio ranges are given for the A stars, WDs, and CVs.}
\tablenotetext{a}{Six are identified based on their $(f_X / f_V)$; three without a cataloged $V$ are identified based on their $(f_X / f_J)$.}
\tablenotetext{b}{Only six WDs have cataloged $V$ magnitudes.}
\tablenotetext{c}{These are log $(f_X / f_B)$ values.}
\end{deluxetable*}

We also show in Figure~\ref{known_ratios} the flux ratio limits for each class as observed by \citet{krautter99}; they identified $274$ stellar X-ray emitters in an area-limited survey of RASS BSC and FSC sources. The \citet{macca88} {\it Einstein} limits are also included in each panel. The flux ratio distributions and ranges for the SIMBAD stars are consistent with those found in these earlier studies. This suggests that, as expected, the vast majority of these stars are RASS source counterparts. Furthermore, the flux ratio ranges calculated from our own distributions for these stellar types give us limits slightly more generous than the {\it Einstein} values \citep[which are known to be quite conservative;][]{stocke91}, but generally more conservative than those of \citet{krautter99}. 

Table~\ref{known_table} summarizes the properties of our sample of SIMBAD stars. We list the total number of stars of each spectral type in the initial sample, the number of known X-ray emitters among them, and the number of new X-ray source counterparts we identify based on the combination of positional proximity and appropriate X-ray-to-optical flux ratios. For the F, G, K, and M stars, we include the empirically derived log $(f_X / f_V)$ means and $2\sigma$ ranges. For the A stars, WDs, and CVs, which are less numerous, we list the full range of flux ratios and their means. For the CVs, we give log ($f_X / f_B$) \citep[see][]{zickgraf03}.

\begin{deluxetable*}{lclcccl}
\tablewidth{0pt}
\tabletypesize{\scriptsize}
\tablecaption{The CVs and WDs among the RASS/SIMBAD matches\label{wd_sim}}
\tablehead{
\colhead{}       & \colhead{}                &\colhead{}            & \colhead{Offset}  & \colhead{$B$}   & \colhead{$V$}   & \colhead{}\\
\colhead{1RXS J} & \colhead{Counts s$^{-1}$} &\colhead{Counterpart} & \colhead{(\asec)} & \colhead{(mag)} & \colhead{(mag)} & \colhead{Type}
}
\startdata
015543.3$+$002817 & $0.042\pm0.012$ & V$*$ FL Cet & $9.9$ & $16.60$ & $15.50$ & P\\
083821.6$+$483800 & $0.565\pm0.042$ & V$*$ EI UMa & $4.2$ & $14.90$ & \nodata & DN\\
085343.5$+$574846 & $0.414\pm0.035$ & V$*$ BZ UMa & $8.0$ & $15.30$ & \nodata & DN\\
113826.8$+$032210 & $0.659\pm0.061$ & V$*$ T Leo  & $2.9$ & $10.00$ & $10.00$ & DN\\
020801.8$+$133625 & $0.054\pm0.016$ & WD 0205$+$133 & $24.7$ & $13.78$ & \nodata & DA+dM1\\
034850.1$-$005823 & $1.965\pm0.158$ & WD 0346$-$011 & $9.2$ & $13.82$ & $13.98$ & DA\\
084104.2$+$032118 & $1.207\pm0.065$ & WD 0838$+$035 & $5.1$ & \nodata & $14.48$ & DA\\
112814.4$-$024950 & $0.051\pm0.015$ & WD 1125$-$025 & $33.1$ & $15.32$ & \nodata & DA\\
152146.6$+$522215 & $0.168\pm0.016$ & WD 1520$+$525 & $11.6$ & $15.56$ & $16.60$ & DO\\
165008.0$+$370130 & $0.115\pm0.019$ & WD 1648$+$371 & $20.3$ & \nodata & $16.10$ & DA\\
165020.4$+$403723 & $0.094\pm0.016$ & WD 1648$+$407 & $3.9$ & \nodata & $14.50$ & DA\\
172642.8$+$583726 & $1.247\pm0.028$ & WD 1725$+$586 & $7.7$ & $15.36$ & $15.45$ & DA 
\enddata
\tablecomments{``P'' indicates that CV is a polar and ``DN'' that it is a dwarf nova.}
\end{deluxetable*}

\begin{figure*}
\centerline{\includegraphics[angle=90,width=1.7\columnwidth]{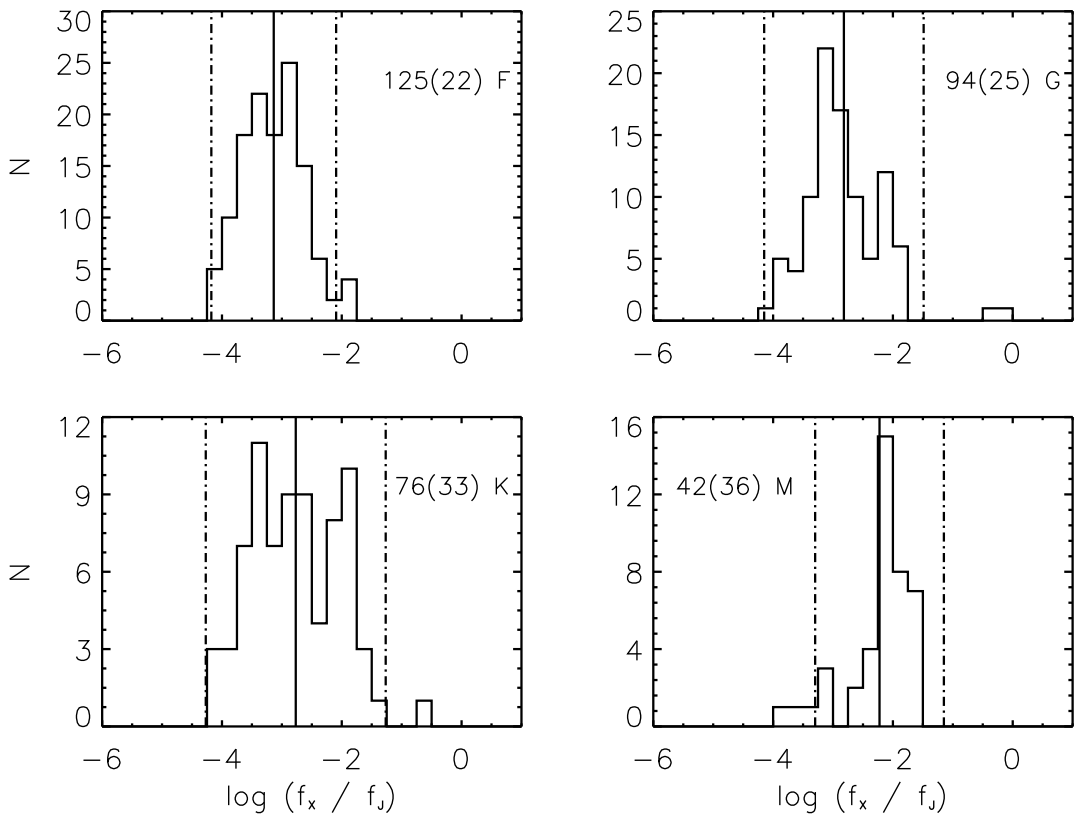}}
\caption{log $(f_X / f_J)$ ratios for the RASS/SIMBAD F, G, K, and M stars. As in Figure~\ref{known_ratios}, the solid vertical lines correspond to the mean $(f_X / f_J)$ and the dot-dashed vertical lines bracket the $2 \sigma$ $(f_X / f_J)$ range for each stellar type. The number of RASS/SIMBAD stars used to calculate these $(f_X / f_J)$ ranges is indicated in each panel; the number in parentheses is the subset of previously known X-ray emitters.}\label{known_J_ratios}
\end{figure*}

Below we touch on the properties of the CVs and WDs in this sample, and discuss our use of the F, G, K, and M stars to construct empirical log $(f_X / f_J)$ ranges\footnote{The A stars are discussed in the Appendix.}.

\subsubsection{The CVs and WDs}\label{wd_cv}
Table~\ref{wd_sim} lists the matched RASS/SIMBAD CVs and WDs. All of the CVs are known X-ray emitters; only EI UMa does not have an SDSS spectrum. \citet{paula1} and \citet{sch2005} identified FL Cet as a polar, where accretion from (typically) a late-type star is directly onto a WD's magnetic poles. The three other CVs are dwarf novae. 

Six of the eight WDs in Table~\ref{wd_sim} are listed as X-ray sources by \citet[][]{fleming96} while another, WD 1520$+$525, was proposed as a soft X-ray source by \citet{chu04}. Seven of these WDs are DAs, stars with pure hydrogen atmospheres that are the most commonly observed WD type. WD 1520$+$525 is the only non-DA; its spectrum includes strong HeII features, making it a DO WD. \citet{wachter03} identified WD 0205$+$133 as belonging to a binary system on the basis of its infrared excess and \citet{fahiri06} resolved its M-star companion; this WD system is the only one not to be previously proposed as an X-ray source. 

Overall, the X-ray sources near the WDs have very soft X-ray spectra, as measured by their hardness ratios. For these sources, HR1 $\approx -1$, which is typical for X-ray-emitting WDs \citep[e.g.,][]{fleming96, zickgraf97}, and confirms that their counterparts are indeed likely to be these WDs. Only one of these objects has an SDSS spectrum: WD 1725$+$586. 

For a broader discussion of RASS detections of WDs and CVs, see \S~\ref{wds_cvs}. 

\subsubsection{Using the known stars to construct empirical log $(f_X / f_J)$ ranges}
While typical flux ratio ranges can be derived using log $(f_X / f_V)$ calculations for our sample of previously cataloged stars, we cannot use this flux ratio to evaluate the likelihood that fainter, previously uncatalogued SDSS stars are X-ray emitters, as they generally lack SIMBAD $V$ magnitudes\footnote{A handful do: a few SIMBAD entries give a magnitude but no spectral class, for example. These stars are included in the APO spectroscopic sample discussed in \S~\ref{spec_sample}.}. While $g$ or $r$ would be a natural substitute for $V$, because many of these stars are saturated, they have unreliable SDSS photometry. Instead, we use 2MASS $J$ magnitudes; all of the stars in our sample have a 2MASS counterpart.

To calculate flux ratios, we adapt the \citet{macca88} formula to reflect the definition of the 2MASS $J$-band flux \citep[e.g.,][]{cohen03}: log $(f_X /f_J)$ = log $(f_X) + 0.4 J + 6.30$. The SIMBAD stars included in our $(f_X / f_J)$ calculations are either previously known X-ray emitters or newly identified RASS counterparts, as defined in \S~\ref{ratio_calc}; in addition, we require that the $J$-band errors $\leq0.11$ mag \citep[for a discussion of the 2MASS flags, see][]{cutri03}. 

\begin{figure*}[t]
\centerline{\includegraphics[width=1.95\columnwidth]{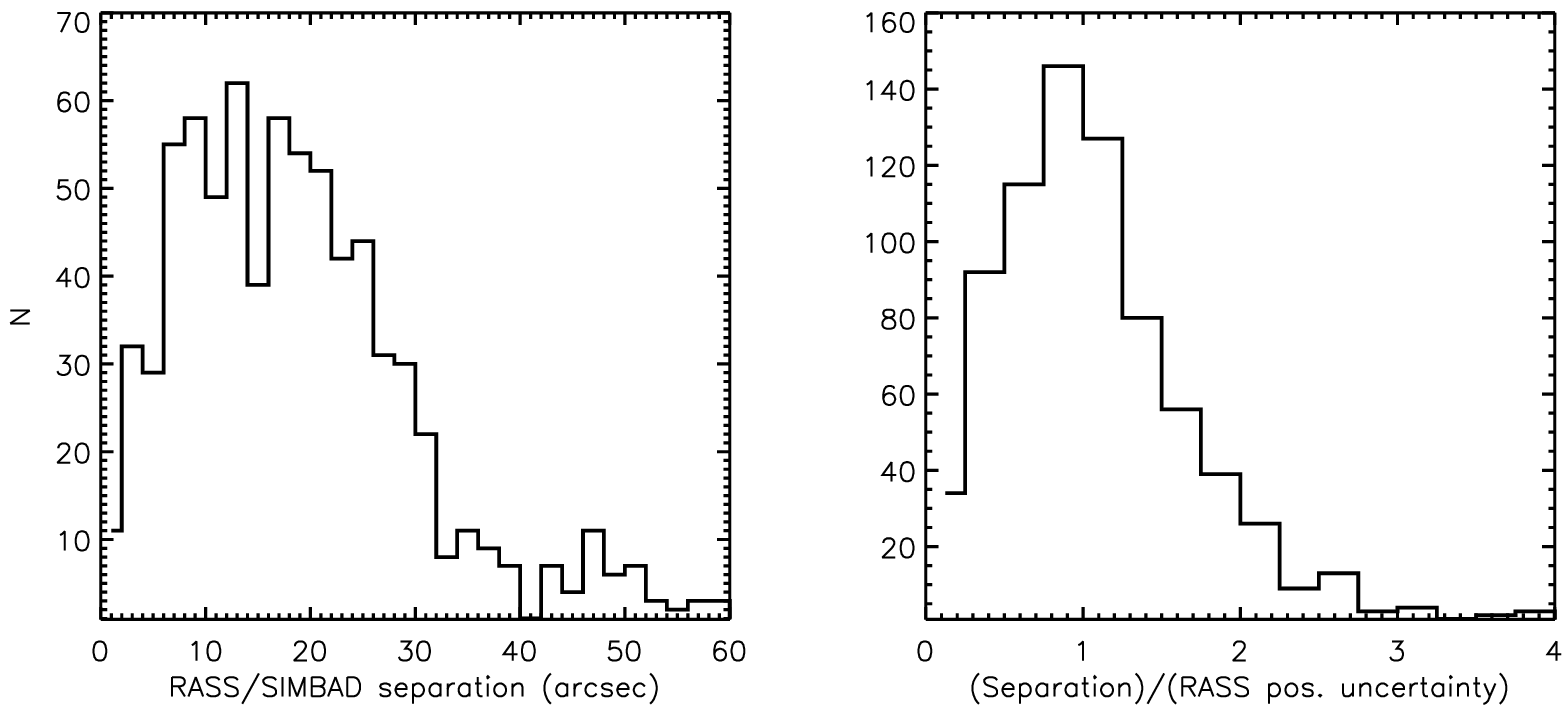}}
\caption{{\it Left panel:} distribution of positional offsets between the RASS sources and $751$ candidate SDSS stellar counterparts with APO spectra. {\it Right panel:} the same distribution, expressed in terms of the cataloged X-ray positional uncertainty.}\label{separation_single}
\end{figure*}

As mentioned above, a small number of these stars have both a cataloged spectral type and one derived from an APO spectrum. In $24$ cases, using either type has no effect on the $(f_X / f_J)$ distributions; in $13$ cases it does. $11$ of these stars were classified as ``F/G'' or ``K'' by \citet{zickgraf03}; these spectra are lower resolution than ours, and we therefore use our APO-derived types instead. The other stars are V* V842 Her and StKM $1-1262$, discussed above; we used the SIMBAD types for these two.

We then have $125$ F, $94$ G, $76$ K, and $42$ M stars with which to construct empirical log $(f_X / f_J)$ ranges. The distributions are shown in Figure \ref{known_J_ratios}. The derived flux ratio $2\sigma$ ranges and means are listed in Table \ref{known_table} alongside the earlier log $(f_X / f_V)$ values. Based on their log $(f_X / f_J)$ values, we are able to add three of the SIMBAD M stars lacking a cataloged $V$ and previously unassociated with RASS sources to our list of new stellar X-ray emitters. In total, we have identified $104$ F, $70$ G, $47$ K, and $9$ M stars cataloged in SIMBAD but previously unassociated with an X-ray source as RASS source counterparts. These are included in our final catalog; see \S~\ref{master_cat}.

\subsection{The APO spectroscopic sample}\label{spec_sample}
Our search for cataloged candidate counterparts identified $\sim1400$ RASS sources with no obvious counterparts in either SIMBAD or NED\footnote{Some sources with matches were included, e.g., matches to Tycho stars lacking a published spectral type.}. The objects $\leq 30''$ from a RASS source made up the list of primary targets for our spectroscopic campaign. We used the Dual Imaging Spectrograph (DIS) on the APO $3.5$-m telescope to obtain spectra for $\sim95\%$ of these objects, and for $>100$ objects with separations $> 30$\asec, over the course of $50$ half nights from 2003 Sep to 2007 Jul. In our standard set-up, the DIS `low' blue grating is centered on $4500$ \AA\ (with a dispersion of $2.4$ \AA\ per pixel); its `medium' red grating is centered on $7600$ \AA\ ($2.3$ \AA\ per pixel); we use a $1.5\asec$ ($3$ pixel) slit. This results in wavelength coverage from $3800$ to $8700$ \AA\ with a resolution of $\sim 700$ in the blue and $\sim 1000$ in the red. Almost all of our observing was done in bright time, and our integrations were typically $\lapprox\ 5$ min. The data were flux calibrated with standards taken each night. The spectra were reduced with a script written in PyRAF, the Python-based command language for the Image Reduction and Analysis Facility (IRAF)\footnote{PyRAF is a product of the Space Telescope Science Institute, which is operated by AURA for NASA. IRAF is distributed by the National Optical Astronomy Observatories, which are operated by the Association of Universities for Research in Astronomy, Inc., under cooperative agreement with the National Science Foundation.}. All spectra were trimmed, overscan and bias corrected, cleaned of cosmic rays, flat fielded, extracted, dispersion corrected, and flux calibrated using standard IRAF tasks.

\begin{figure*}
\centerline{\includegraphics[angle=90,width=1.7\columnwidth]{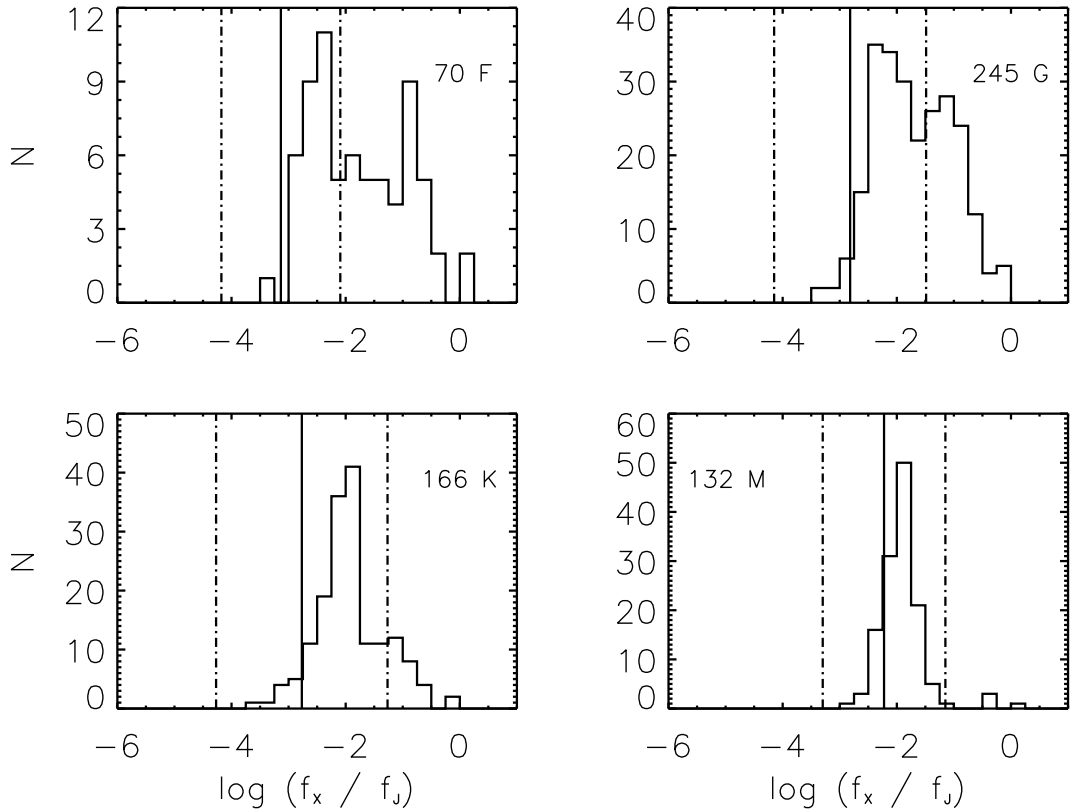}}
\caption{log $(f_X / f_J)$ ratios for the candidate SDSS counterparts with APO spectroscopy. The solid vertical lines correspond to the mean $(f_X / f_J)$ and the dot-dashed vertical lines bracket the $2 \sigma$ $(f_X / f_J)$ range for each stellar type as derived from the RASS/SIMBAD sample. The number of SDSS/APO stars used to calculate these $(f_X / f_J)$ ranges is indicated in each panel.}\label{ratio_all_stars}
\end{figure*}

In most cases, a single observation was sufficient to obtain a spectral type for the candidate counterpart; we thus have $649$ error circles with a unique spectrum. In $107$ error circles we took at least two spectra of the same target, generally because the target was relatively faint ($g\ \gapprox\ 16$) or the weather uncooperative, and the first spectrum was of insufficient quality to allow for confident typing. In an additional $98$ RASS error circles we took spectra for more than one target, usually because of the presence of an apparent companion to our target. The total number of RASS sources considered is nearly $900$ if we include the $41$ SIMBAD stars with cataloged spectral types for which we obtained APO spectra.

\begin{deluxetable*}{lclcccccl}
\tabletypesize{\scriptsize}
\tablewidth{0in}
\setlength{\tabcolsep}{-1in}
\tablecaption{RASS sources with multiple candidate counterparts\label{multiple_m}}
\tablehead{
\colhead{      } & \colhead{               } & \colhead{} & \colhead{Offset  }   & \colhead{$g$}   & \colhead{$J$}   & \colhead{log}         & \colhead{APO} & \colhead{SIMBAD} \\
\colhead{1RXS J} & \colhead{Counts s$^{-1}$} & \colhead{SDSS J} & \colhead{($\asec$)}  & \colhead{(mag)} & \colhead{(mag)} & \colhead{$(f_X/f_J)$} & \colhead{type} & \colhead{data}
}
\startdata
002203.7$-$... & $0.135\pm0.024$ & 002204.47$-$... & $14.5$ & $12.98\pm0.01$ & $9.51\pm0.02$  & $-1.77$ & K7e & \nodata \\
                  &               & 002204.20$-$... & $16.5$ & $14.55\pm0.05$ & $10.00\pm0.02$ & $-1.57$ & K5 & \nodata \\	
083203.9$+$... & $0.020\pm0.010$ & 083203.74$+$... & $5.8$  & $12.33\pm0.02$ & $10.08\pm0.02$ & $-2.37$ & K1 & \nodata \\ 	
                  &               & 083202.66$+$... & $12.7$ & $13.48\pm0.13$ & $10.27\pm0.02$ & $-2.30$ & G0 & \nodata \\ 	
092455.1$+$... & $0.130\pm0.023$ & 092456.44$+$... & $11.2$ & $14.09\pm0.02$ & $9.83\pm0.02$  & $-1.66$ & M3 & \nodata \\
                  &               & 092455.33$+$... & $10.7$ & $14.62\pm0.08$ & $8.66\pm0.02$  & $-2.12$ & M2 & RX J0924.9$+$5756 1; $*$ \\
110830.8$+$... & $0.073\pm0.015$ & 110830.87$+$... & $1.4$  & $11.92\pm0.00$ & $10.28\pm0.03$ & $-1.73$ & G7 & \nodata \\
                  &               & 110830.87$+$... & $1.4$  & $11.92\pm0.00$ & $10.35\pm0.07$ & $-1.70$ & G2 & RX J1108.5$+$0117 \\
114930.0$-$... & $0.052\pm0.016$ & 114930.34$-$... & $10.8$ & $11.37\pm0.00$ & $9.53\pm0.03$  & $-2.17$ & G0 & \nodata \\	
                  &               & 114930.34$-$... & $10.8$ & $11.37\pm0.00$ & $9.53\pm0.03$  & $-2.17$ & G6 & \nodata \\	
123247.7$-$... & $0.024\pm0.010$ & 123247.68$-$... & $0.4$  & $14.38\pm0.20$ & $10.11\pm0.03$ & $-2.28$ & G6 & \nodata \\	
                  &               & 123247.28$-$... & $7.5$  & $15.37\pm0.15$ & $10.69\pm0.03$ & $-2.05$ & G8 & \nodata \\	
132339.1$+$... & $0.021\pm0.009$ & 132338.98$+$... & $4.8$  & $15.10\pm0.07$ & $11.4\pm0.03$  & $-1.81$ & M0 & \nodata \\	
                  &               & 132339.64$+$... & $5.8$  & $13.90\pm0.03$ & $10.67\pm0.05$ & $-2.11$ & M1 & \nodata \\	
135902.2$+$... & $0.067\pm0.013$ & 135902.75$+$... & $10.4$ & $12.32\pm0.00$ & $9.42\pm0.03$  & $-2.11$ & K5 & GPM 209.761906$+$...; $11.30$; $-2.11$; $*$\\
                  &               & 135903.24$+$... & $10.3$ & $16.21\pm0.07$ & $11.52\pm0.03$ & $-1.27$ & M3 & \nodata \\	
140148.9$+$... & $0.024\pm0.011$ & 140148.48$+$... & $17.9$ & $14.35\pm0.11$ & $11.03\pm0.03$ & $-1.90$ & K4 & \nodata \\	
                  &               & 140148.69$+$... & $17.8$ & $14.42\pm0.02$ & $10.75\pm0.06$ & $-2.02$ & M0 & \nodata \\	
150656.8$+$... & $0.025\pm0.010$ & 150656.87$+$... & $1.4$  & $15.37\pm0.02$ & $10.84\pm0.05$ & $-1.97$ & M3 & \nodata \\	
                  &               & 150656.90$+$... & $6.6$  & $16.65\pm0.12$ & $11.31\pm0.03$ & $-1.78$ & M3 & \nodata \\	
153826.3$+$... & $0.010\pm0.004$ & 153826.02$+$... & $7.5$  & $17.88\pm14.96$ & $9.11\pm0.04$ & $-3.06$ & K3 & BD$+$53 1797B; $10.70$; $-3.06$; $*$i$*$ \\ 
                  &               & 153826.38$+$... & $11.3$ & $14.38\pm0.16$ & $8.88\pm0.03$  & $-3.15$ & K3 & HD 234250; $9.50$; $-3.15$; K2; $*$i$*$ \\
164129.8$+$... & $0.022\pm0.009$ & 164129.54$+$... & $17.7$ & $20.02\pm13.33$ & $8.92\pm0.04$ & $-2.79$ & G8 & BD$+$40 3051\tablenotemark{a}; $*$ \\ 
                  &               & 164129.47$+$... & $14.5$ & $10.15\pm0.00$ & $8.92\pm0.04$  & $-2.79$ & G5 & \nodata \\	
171017.5$+$... & $0.052\pm0.006$ & 171017.05$+$... & $6.8$  & $13.23\pm0.07$ & $9.77\pm0.03$  & $-2.08$ & K4 & 2E 1709.9$+$6325 \\
                  &               & 171016.46$+$... & $7.1$  & $19.96\pm6.98$ & $10.05\pm0.02$ & $-1.96$ & M0 & 2E 1709.9$+$6325 \\
235942.7$-$... & $0.029\pm0.011$ & 235943.68$-$... & $15.0$ & $15.85\pm0.11$ & $9.94\pm0.05$  & $-2.26$ & K7 & \nodata \\	
                  &               & 235943.80$-$... & $16.6$ & $14.18\pm0.03$ & $10.47\pm0.03$ & $-2.05$ & K3 & \nodata \\	
\tableline
001619.1$-$... & $0.020\pm0.009$ & 001620.75$-$... & $25.5$ & $14.85\pm0.03$ & $12.57\pm0.03$ & $-1.37$ & K3e & GSC 04664$-$01105; $*$ \\
                  &               & 001620.16$-$... & $17.6$ & $15.51\pm0.03$ & $12.27\pm0.02$ & $-1.49$ & K7 & \nodata \\	
015119.8$+$... & $0.172\pm0.023$ & 015119.97$+$... & $13.2$ & $11.86\pm0.00$ & $9.08\pm0.02$  & $-1.83$ & G9 & [ZEH2003] RX J0151.3$+$1324 2 \\ 
                  &               & 015119.97$+$... & $7.7$  & $14.85\pm0.08$ & $8.56\pm0.03$  & $-2.04$ & M3 & \nodata \\	
020419.0$+$... & $0.053\pm0.016$ & 020419.12$+$... & $5.4$  & $18.37\pm92.46$ & $9.98\pm0.02$ & $-1.99$ & M3 & [ZEH2003] RX J0204.3$+$1318 2 \\
                  &               & 020418.31$+$... & $10.3$ & $16.00\pm0.01$ & $11.33\pm0.02$ & $-1.45$ & K0 & \nodata \\	
085921.0$+$... & $0.022\pm0.009$ & 085922.99$+$... & $22.3$ & $12.63\pm0.00$ & $10.85\pm0.03$ & $-2.03$ & G8 & \nodata \\	
                  &               & 085922.79$+$... & $23.1$ & $16.61\pm0.24$ & $12.92\pm0.05$ & $-1.20$ & M0 & \nodata \\	
091140.6$+$... & $0.018\pm0.008$ & 091139.52$+$... & $12.7$ & $17.63\pm0.02$ & $12.77\pm0.02$ & $-1.34$ & M4 & \nodata \\	
                  &               & 091140.92$+$... & $7.1$  & $15.36\pm0.02$ & $11.67\pm0.02$ & $-1.78$ & M1 & \nodata \\	
145626.1$+$... & $0.019\pm0.007$ & 145628.50$+$... & $28.0$ & $10.96\pm0.00$ & $9.60\pm0.02$  & $-2.57$ & G0 & TYC 3864$-$410$-$1; $10.63$; $-2.57$; $*$ \\
                  &               & 145626.46$+$... & $17.2$ & $13.55\pm0.01$ & $10.69\pm0.02$ & $-2.13$ & K5 & \nodata \\	
150521.8$+$... & $0.010\pm0.004$ & 150523.09$+$... & $18.0$ & $15.69\pm0.18$ & $10.01\pm0.02$ & $-2.70$ & G6 & \nodata \\	
                  &               & 150520.38$+$... & $18.5$ & $11.25\pm0.00$ & $9.83\pm0.02$  & $-2.78$ & G6 & \nodata \\	
152350.6$+$... & $0.018\pm0.007$ & 152348.89$+$... & $18.3$ & $13.72\pm0.01$ & $9.33\pm0.02$  & $-2.72$ & M3 & NLTT 40162; PM$*$ \\
                  &               & 152351.61$+$... & $28.9$ & $18.91\pm0.02$ & \nodata        & $+0.19\tablenotemark{b}$  & QSO & \nodata \\	
160207.9$+$... & $0.013\pm0.006$ & 160206.84$+$... & $27.2$ & $12.25\pm0.00$ & $7.71\pm0.02$  & $-3.49$ & G9 & TYC 3877$-$1522$-$1; $9.61$; $-3.49$; $*$ \\
                  &               & 160208.39$+$... & $17.6$ & $13.84\pm0.02$ & $11.47\pm0.02$ & $-1.99$ & K1 & \nodata \\	
163845.2$+$... & $0.022\pm0.007$ & 163845.73$+$... & $7.4$  & $12.41\pm0.00$ & $10.45\pm0.02$ & $-2.18$ & K2 & \nodata \\	
                  &               & 163844.64$+$... & $17.2$ & $15.36\pm0.01$ & $12.27\pm0.02$ & $-1.46$ & K7 & \nodata \\	
171702.0$+$... & $0.047\pm0.007$ & 171702.55$+$... & $15.8$ & $13.40\pm0.02$ & $9.52\pm0.02$ & $-2.22$ & G6 & \nodata \\	
                  &               & 171702.12$+$... & $1.6$ & $13.99\pm0.05$ & $8.62\pm0.04$ & $-2.58$ &  K2 & TYC 3891$-$410$-$1; $10.24$; $-2.58$; $*$ \\ 
172713.4$+$... & $0.013\pm0.004$ & 172715.09$+$... & $16.0$ & $15.42\pm0.02$ & $11.3\pm0.02$ & $-2.08$ & G7 & \nodata \\	
                  &               & 172714.58$+$... & $16.7$ & $10.63\pm0.00$ & $8.97\pm0.02$ & $-3.02$ & G8 & TYC 3900$-$417$-$1; $10.30$; $-3.02$; $*$ 
\enddata
\tablecomments{Pairs above the line are close companions. In several cases, the SDSS pipeline returns the same data for both stars. The SIMBAD data includes, when available, the cataloged object name, its $V$ magnitude, the corresponding log $(f_X/f_V)$, and/or its type. ``$**$'' or ``$*$i$*$'' indicates a double or multiple star system and ``PM$*$'' a high proper-motion star.}
\tablenotetext{a}{The given SIMBAD position is offset by $37\asec$, but this appears to be the SDSS star.} 
\tablenotetext{b}{log $(f_X/f_V)$. There is no 2MASS detection of this QSO.} 
\end{deluxetable*}

To obtain spectral types for our targets, we used the Hammer \citep{kev2007}. The Hammer automatically predicts the Morgan-Keenan (for stars earlier than M) or Kirkpatrick (for later stars) spectral type on the basis of a fit to a set of $30$ spectral indices\footnote{The Hammer is written in IDL and is available from {\tt http://www.cfa.harvard.edu/$\sim$kcovey/}.}. The Hammer also allows the user to interactively modify the assigned spectral type. Every spectrum was therefore checked by eye before a final type was assigned.

As noted above, in $107$ cases we obtained at least two spectra of a candidate RASS counterpart, generally because our first spectrum was fairly noisy. Nevertheless, $80\%$ of the time the stellar types assigned to a given star agree to within two subclasses, which we take to be our uncertainty. For those cases where the difference is larger, we assigned the spectral type obtained from fitting the highest signal-to-noise spectrum for that particular star.

In Figure~\ref{separation_single} we present the positional offset and $\sigma$ (as defined earlier) distributions for stars in $751$ of the $756$ RASS error circles where we selected only one target. For these calculations, we updated the positional (and photometric) data for the SDSS stars to the more recent Data Release 6 \citep[DR6;][]{DR6paper} values. Of the five objects not included, one is a QSO, two are stars whose spectra are too poor for confident typing, and two others lack data in DR6 because they fall on the survey edges. 

$646$ of the $751$ stars have separations $\leq 30$\asec; to these we add two stars with separations $> 30$\asec\ previously identified as RASS counterparts by \citet{zickgraf03} but lacking a spectral type. Of these $648$ stars, $645$ are F, G, K, or M stars (two others were typed as A stars\footnote{Discussed in the Appendix.} and one is a CV). If we apply the same $J$-band quality cut as to our earlier sample ($J_{err} \leq 0.11$ mag), and remove objects for which the RASS flag is not $0$, we are left with a sample of $616$ stars: $70$ F, $245$ G, $166$ K, and $132$ M. 

The log $(f_X / f_J)$ distributions for these F, G, K, and M stars are plotted in Figure~\ref{ratio_all_stars}. We also include the flux ratio means and $2 \sigma$ ranges derived from the RASS/SIMBAD sample (see Table~\ref{known_table}). We identify as spectroscopically confirmed counterparts to RASS X-ray sources objects with $(f_X / f_J)$ within $2 \sigma$ of the mean $(f_X / f_J)$ value of optically bright X-ray emitters in the same spectral type range. These plots show clearly that the probability that a spectroscopic target is a RASS counterpart increases as we move to later spectral types. Only $31$ of the $70$ spectroscopically identified F stars fall within the $2 \sigma$ log $(f_X / f_J)$ range derived from the RASS/SIMBAD F stars ($44\%$); for the G stars, the fraction is $149$ of $245$ ($61\%$). For K stars and M stars the fraction is much higher: $139$ of $166$ K stars ($84\%$) and $127$ of $132$ M stars ($96\%$) have log $(f_X / f_J)$ within their respective $2 \sigma$ ranges. 

The X-ray, optical, and infrared properties of the $443$ new F, G, K, and M X-ray emitters identified here are included in our final catalog\footnote{Note that this number reflects our removal of three stars (one G, one K, and one M) observed at APO from our final catalog based on our comparison to the \citet{flesch04} catalog. See \S~\ref{fpcat}.}, described in \S~\ref{master_cat}, along with the $230$ RASS/SIMBAD F, G, K, and M stars identified above. Below we discuss the error circles for which we obtained multiple APO spectra, our candidate flare stars, and a new CV discovered in this spectroscopic campaign. 

\subsubsection{The error circles with multiple spectra}
For the error circles in which we have APO spectra for several targets, we calculated the positional separation and log $(f_X/f_J)$ ratio and obtained a spectral type for each. The best candidate counterpart in each error circle was then identified on the basis of its proximity (only stars $\leq 30$\asec\ from a RASS source whose X-ray flag is $0$ were considered) and the appropriateness of its $(f_X/f_J)$ given its spectral type. In $34$ error circles we identified a single stellar counterpart ($2$ F, $11$ G, $8$ K, and $13$ M stars); these are included in our final catalog. 

In another $26$ error circles, two stars met our criteria; in $14$ of these cases our target has an apparent close companion (separations of $\sim 5''$ or less), making it difficult to avoid photometric uncertainties in the SDSS and 2MASS catalogs and sometimes to obtain clean spectra with our $1.5''$ slit. We list these error circles separately in Table~\ref{multiple_m}. Higher resolution X-ray observations are required to determine which of the possible counterparts is in fact the X-ray source; it is also possible, of course, that both stars are X-ray emitters.

\begin{figure}
\epsscale{1.2}
\centerline{\plotone{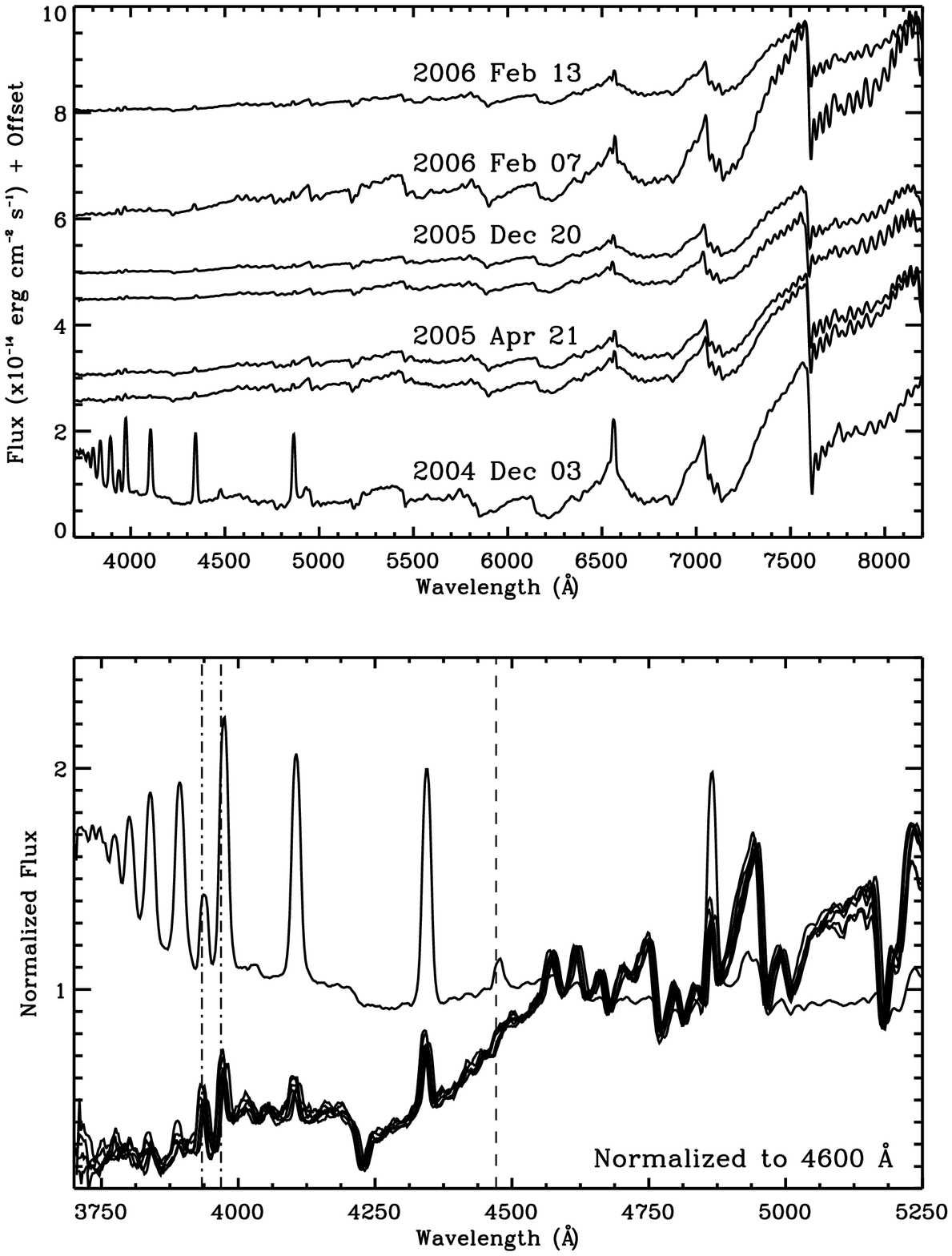}}
\caption{APO spectra of the proposed stellar counterpart to 1RXS J080826.7$+$434745. {\it Top}: the spectra are artificially offset from each other in flux; two spectra were obtained on 2005 April 21 and December 20. {\it Bottom}: the blue continuum and strong Balmer emission visible in the 2004 December 03 spectrum are even more apparent when all of the spectra are normalized to their flux value at $4600$ \AA. The dot-dashed lines indicate the positions of the Ca II H \& K lines (at this resolution, Ca H is blended with H$\epsilon$), while the dashed line is HeI ($4471$ \AA).}\label{0808_spec}
\end{figure}

\subsubsection{The candidate flare stars}\label{flares}
Many of the stars for which we obtained multiple spectra were M stars with what appeared to be stronger-than-average emission lines. Figure~\ref{0808_spec} shows the spectra for the target with the largest number of observations, the proposed counterpart to 1RXS J080826.7$+$434745. The initial spectrum suggested that we caught this M4 star flaring, and we returned to it regularly over the course of the next two years. All of the spectra include the Balmer and CaII H \& K emission lines typical of M star spectra. However, these features are much stronger in the initial spectrum, as can be clearly seen in the bottom panel of Figure~\ref{0808_spec}, where the spectra have been normalized to the value of their flux at $4600$ \AA. In addition, an emission feature at $4471$ \AA\ is evident in the initial spectrum and absent in the quiescent spectra. During a flare, the amount of material emitting at chromospheric temperatures increases, and lines due to high-excitation species such as HeI, which is responsible for this feature, appear \citep{nlds}. The blue continuum visible in the initial spectrum is also typically observed in flare events \citep[e.g.,][]{hawley91}. In a forthcoming paper (E.\ Hilton et al., 2009, in preparation) we describe the properties of M stars for which we detected similar changes to the emission lines and continuum.

\begin{figure}[!h]
\epsscale{1.2}
\centerline{\plotone{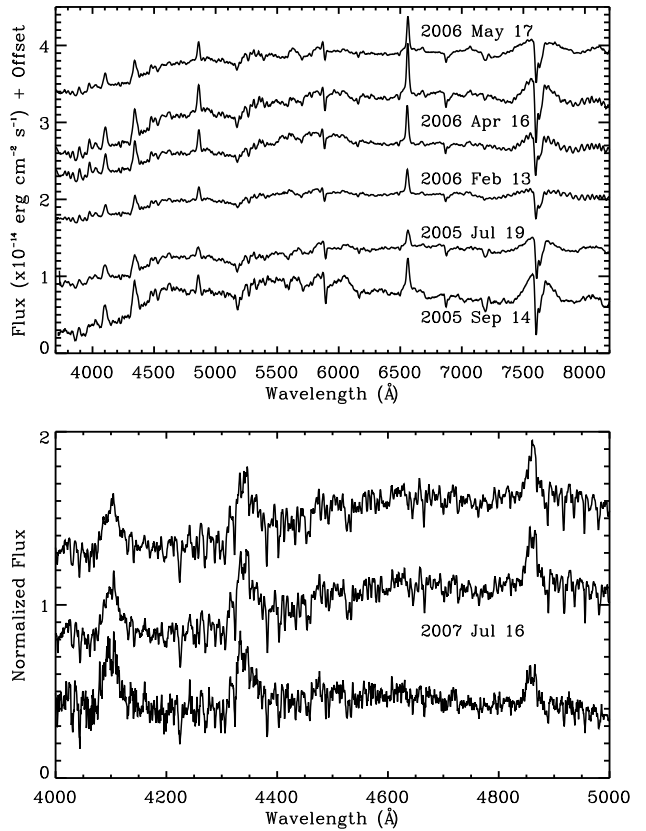}}
\caption{APO spectra of the proposed stellar counterpart to 1RXS J171456.2$+$585130. {\it Top:} low-resolution spectra, presented as in Figure~\ref{0808_spec}. Two spectra were obtained on 2006 Apr 16. {\it Bottom:} high-resolution spectra of the region including the H$\beta$, H$\gamma$, and H$\delta$ lines showing the structure typically seen in CVs.}\label{1714_spec}
\end{figure}

\subsubsection{A new CV: SDSS J171456.78$+$585128.3}
APO spectra for the proposed counterpart to the RASS source RXS J171456.2$+$585130 are presented in Figure~\ref{1714_spec}. We obtained six spectra for this star on five separate nights between 2005 Sep 14 and 2006 May 17 using DIS in the low-resolution setup. While the underlying spectrum corresponds to that of a K4 star, strong, persistent Balmer emission is obvious (see the top panel of Figure~\ref{1714_spec}). The emission lines are broader than for the M star in Figure~\ref{0808_spec}, suggesting that we are observing accretion rather than flaring, and there is therefore a currently invisible companion to the K4 star.

On 2007 Jul 16, we obtained three more spectra of this star using DIS in the high-resolution setup ($0.62$ \AA\ per pixel in the blue, $0.56$ \AA\ in the red, with the gratings centered on $4600$ and $6800$ \AA\ respectively). These spectra confirmed that the emission lines are broader than those of flaring stars, and also more structured than those typically seen in flares or in symbiotic systems (see bottom panel of Figure~\ref{1714_spec}). We used IRAF's {\it splot} routine to measure the H$\alpha$ and H$\beta$ equivalent widths (EqWs) in our two longest observations ($20$ and $30$ min). The average H$\alpha$ EqW $= 12.5$ \AA, corresponding to a flux of $ 9.5 \times 10^{-14}$ ergs cm$^{-2}$ s$^{-1}$; for H$\beta$, the EqW $=10.0$ \AA\ and the flux $= 7.0 \times 10^{-14}$ ergs cm$^{-2}$ s$^{-1}$. We conclude that SDSS J1714 is a new CV.

To measure its period, we used DIS in its new low-resolution setup ($1.9$ \AA\ per pixel in the blue, $2.3$ \AA\ in the red, with the gratings centered on $4295$ and $6600$ \AA) to acquire a dozen $20$-min spectra on 2007 Oct 07. Since the secondary in this system is a K star, we expected the orbital period to be long, on the order of 5 to 6 hours. In fact, our radial velocity measurements suggest that the period is $\sim10$ hours and further observations are required to determine SDSS J1714's period and nature. Its log $(f_X/f_g)$ of $-1.44$, while low, is consistent with it being a dwarf nova, as is its position in the log $(f_X/f_g)$ vs.\ H$\beta$ EqW plot (see \S~\ref{cv_details}). 

\section{The Catalog}\label{master_cat}
\subsection{Distances and X-ray Luminosities}
We derive distances to the F, G, K, and M stars in our catalog using photometric parallax relations appropriate for dwarfs on the main sequence, since these dominate our sample. We construct $B-K_s$ colors using the Tycho/USNO $B$ magnitudes for $662$ stars in the catalog ($48$ lack a $B$ magnitude) and adopt the relationship between $g-K_s$ and $B-K_s$ derived by \citet{us2008}. We then use a fit to the M$_{K_s}$ vs.\ $g-K_s$ tabulations of \citet{Kraus2007} to derive distances to each star. Adopting the same conversion of $1$ count s$^{-1}$ into $f_X = 10^{-11}$ erg cm$^{-2}$ s$^{-1}$ used in \S~\ref{ratio_calc}, we calculate  X-ray luminosities for our stars. 

\begin{figure}
\epsscale{1.1}
\centerline{\plotone{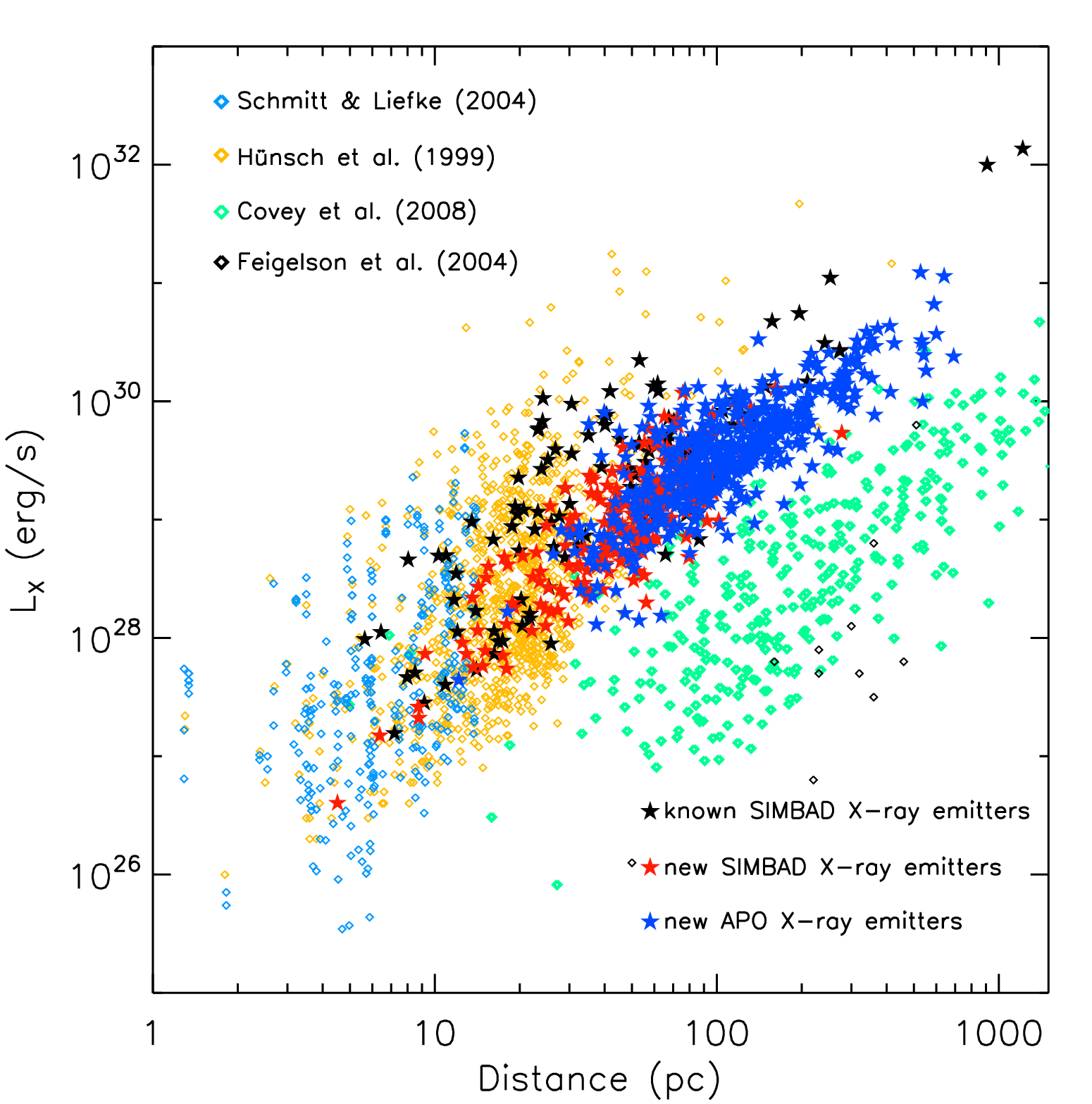}}
\caption {$L_X$ as a function of distance for our sample of X-ray emitting stars. Also shown are the samples of \citet{schmitt04}, \citet{hunsch99}, \citet{us2008}, and \citet{feigelson04}.}\label{lxvsdist}
\end{figure}

These luminosities are shown in Figure~\ref{lxvsdist} as a function of distance. We also plot data from several other catalogs of X-ray emitters. The \citet{schmitt04} and \citet{hunsch99} catalogs are both of {\it ROSAT} detections of nearby stars, and the luminosities are also over the $0.1 - 2.4$ keV range. We also include the $11$ X-ray emitting stars identified by \citet{feigelson04} in the {\it Chandra} Deep Field North and the $348$ identified by \citet{us2008} in the Extended {\it Chandra} Multiwavelength Project (here we use the $0.5-2.0$ keV luminosities). Our sample covers a unique area in the $L_X$--distance plane, with $10^{27}\ \lapprox\ L_X\ \lapprox\ 10^{31}$ ergs s$^{-1}$ and $10\ \lapprox\ $D $\lapprox\ 1000$ pc. 

\begin{deluxetable*}{lclccccccc}[!t]
\tabletypesize{\scriptsize}
\setlength{\tabcolsep}{0.01in}
\tablecaption{Sample data and derived quantities for the stars in our catalog.\label{master_table}}
\tablehead{
\colhead{}       & \colhead{}                & \colhead{} 
      & \colhead{Offset}    & \colhead{SDSS $g$}      & \colhead{2MASS $J$}   & 
\colhead{log}         & \colhead{Syn.\ M$_{K_s}$} & \colhead{Distance} & \colhead{L$_X$} \\
\colhead{1RXS J} & \colhead{Counts s$^{-1}$} & \colhead{SDSS J} & \colhead{$(\asec)$} & \colhead{(mag)} & \colhead{(mag)} & \colhead{$(f_X/f_J)$} & \colhead{(mag)}   & \colhead{(pc)}    & \colhead{(ergs s$^{-1}$)} 
}
\startdata
000053.1$+$... & $0.025\pm0.011$ & 000051.96$+$... & $20.5$ & $12.79\pm0.00$ & $10.30\pm0.02$ & $-2.18$ & $4.63\pm0.08$ & $96\pm4$ & $2.8\pm1.2\times10^{29}$ \\
000312.8$-$... & $0.025\pm0.011$ & 000312.63$-$... & $20.2$ & $14.37\pm0.02$ & $11.30\pm0.02$ & $-1.78$ & $4.72\pm0.09$ & $146\pm6$ & $6.4\pm2.7\times10^{29}$ \\
000447.9$-$... & $0.051\pm0.015$ & 000446.97$-$... & $19.0$ & $11.66\pm0.00$ & $8.54\pm0.03$ & $-2.58$ & $4.45\pm0.07$ & $52\pm2$ & $1.6\pm0.5\times10^{29}$ \\
000659.3$-$... & $0.023\pm0.009$ & 000658.40$-$... & $15.4$ & $12.59\pm0.00$ & $10.39\pm0.03$ & $-2.18$ & $4.63\pm0.08$ & $106\pm4$ & $3.1\pm1.2\times10^{29}$ \\
000714.2$-$.. & $0.062\pm0.016$ & 000714.06$-$... & $5.6$ & $12.60\pm0.00$ & $8.83\pm0.03$ & $-2.38$ & $5.89\pm0.16$ & $26\pm2$ & $5.1\pm1.3\times10^{28}$ \\
000803.1$-$... & $0.041\pm0.012$ & 000803.42$-$... & $6.6$ & $15.92\pm0.02$ & $10.91\pm0.02$ & $-1.73$ & $7.17\pm0.36$ & $38\pm6$ & $7.1\pm2.2\times10^{28}$ \\
000921.2$+$... & $0.114\pm0.019$ & 000921.79$+$... & $9.4$ & $12.09\pm0.00$ & $9.44\pm0.02$ & $-1.87$ & $4.68\pm0.08$ & $64\pm3$ & $5.6\pm0.9\times10^{29}$ \\
001745.3$-$... & $0.097\pm0.019$ & 001745.03$-$... & $5.0$ & $10.34\pm0.00$ & $8.68\pm0.03$ & $-2.24$ & $3.94\pm0.18$ & $73\pm6$ & $6.2\pm1.2\times10^{29}$ \\
002127.8$-$... & $0.057\pm0.017$ & 002127.08$-$... & $15.5$ & $12.63\pm0.00$ & $10.53\pm0.03$ & $-1.73$ & $4.45\pm0.07$ & $127\pm4$ & $1.1\pm0.3\times10^{30}$ \\
002552.2$-$... & $0.055\pm0.016$ & 002550.97$-$... & $24.4$ & $14.80\pm0.03$ & $9.88\pm0.03$ & $-2.01$ & \nodata & \nodata & \nodata 
\enddata
\tablecomments{This table is available in its entirety in a machine-readable form in the online journal. A portion is shown here for guidance regarding its form and content.}
\end{deluxetable*}

\subsection{Comparison to the \citet{flesch04} and \citet{parejko08} catalogs}\label{fpcat}
As a test of the robustness of our identifications, we queried the \citet{flesch04} catalog for matches to our own. They provide classifications for $115$ X-ray sources that appear in our catalog\footnote{In nine cases there are two entries in the \citet{flesch04} catalog for what appear to be different detections of the same X-ray source.}. $100$ have counterparts that are classified as stars; while these are therefore previously known sources, our work provides additional information (e.g., spectral types) not available in the \citet{flesch04} catalog.

The $15$ other sources have counterparts identified by \citet{flesch04} as radio sources (six), galaxies (nine), or AGN/QSOs (three). However, in four of these cases, \citet{flesch04} calculate that their catalog-level matching has most likely produced a random association between e.g., a radio source and an optical object. For five sources, inspection of the SDSS data for the RASS error circles reveals that the position of the \citet{flesch04} counterpart is that of our proposed stellar counterpart, or is offset but still consistent with being the star in our catalog (i.e., there is no SDSS object detected at the \cite{flesch04} position and the nearest object is the bright star we identify as the source). The available SDSS data indicate that the proposed AGN/QSOs are the RASS source counterparts, and we remove these three sources from our catalog. This leaves only three cases in which \citet{flesch04} propose a different (non-AGN) counterpart to the X-ray source.

In another $171$ cases a source in our catalog is classified by \citet{flesch04} simply as an X-ray source. They calculate that $39$ of these entries most probably have stellar counterparts and that for another $50$ sources their proposed association is most likely random. Inspection of the SDSS data for the remaining $82$ \citet{flesch04} RASS error circles finds that in $60$ cases the counterpart's position is that of our cataloged star, or is offset but still consistent with being the star in our catalog. For $149$ of these $171$ sources, therefore, the \citet{flesch04} classification either agrees with or does not contradict our own. 

We also checked our catalog for entries listed in the \citet{parejko08} catalog of $\sim1900$ X-ray emitting galaxies, constructed from matching the RASS to the so-called main galaxy sample from the SDSS Data Release 4 \citep[DR4;][]{DR4paper}. $30$ sources in our catalog are listed as having an SDSS galaxy as the X-ray counterpart. This sample is dominated by unclassifiable galaxies ($24$), which \citet{parejko08} define as galaxies for which most or all of their chosen diagnostic emission lines have fluxes too weak to be measured, and for which most of the matches to the RASS are in fact random.  Unsurprisingly, only two of these galaxies are quoted as having a more than $50\%$ chance of being the X-ray counterpart; $17$ are less than $10\%$ likely to be the RASS source. Of the six remaining galaxies, only a transition galaxy and a galaxy with unclassified emission, meaning that its position in the standard BPT diagrams \citep[][]{baldwin81} shifts according to the line ratios being considered, have a more than $50\%$ chance of being the X-ray source counterpart.

Given this broad agreement between our classifications and those of both \citet{flesch04} and \citet{parejko08} and the additional information provided by our catalog, we list all of the (non-AGN) sources that appear in their catalogs and ours in our final catalog and note their \citet{flesch04} and/or \citet{parejko08} classification. 

\subsection{Available data}
The full version of our catalog of $709$ stars is available on-line only. Sample data and derived quantities are shown in Table~\ref{master_table}. We give the RASS source name, positional error, and count rate, the SDSS counterpart's name, the offset between the X-ray and optical position, and the SDSS $g$ and 2MASS $J$ magnitudes. We caution that any SDSS photometry for objects brighter than $g \sim 14$ is suspect, and should be used only once the objects' flags (included in the full on-line catalog) have been considered. We then provide the calculated log $(f_X/f_J)$, the synthetic M$_{K_s}$ and corresponding distance for stars with a cataloged $B$, and the resulting X-ray luminosity and associated error (assuming $1$ count~s$^{-1}$ corresponds to $f_X = 10^{-11}$~erg~cm$^{-2}$~s$^{-1}$).

The on-line catalog contains the complete SDSS and 2MASS photometry, and where relevant, SIMBAD data or APO observation dates and derived spectral types, as well as classification information from the \citet{flesch04} or \citet{parejko08} catalog.

\begin{figure*}
\centerline{\includegraphics[angle=90,width=1.7\columnwidth]{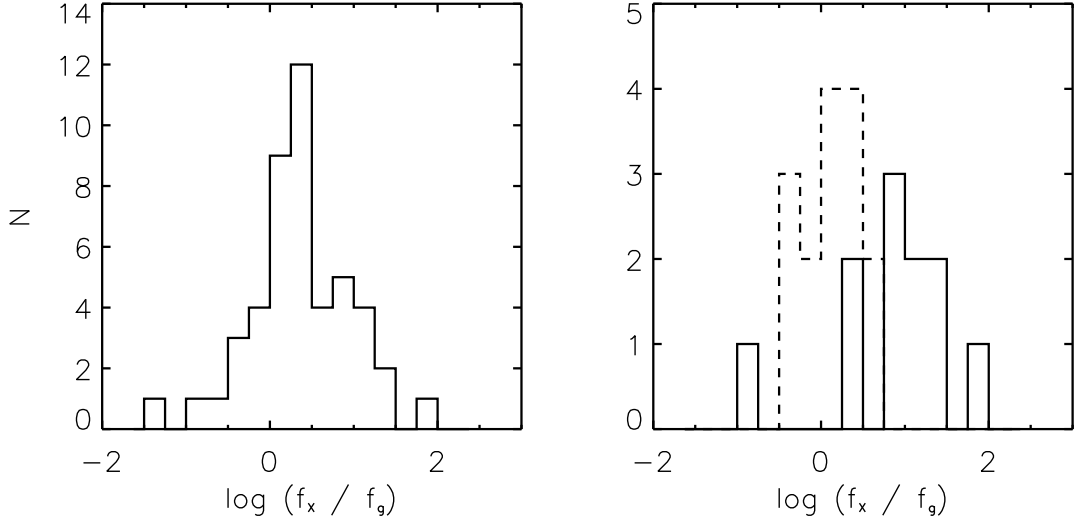}}
\caption{{\it Left panel:} distribution of log $(f_X/f_g)$ for $47$ SDSS CVs identified as RASS counterparts. {\it Right panel:} the same distribution for the known dwarf novae (dashed histogram) and polars (solid histogram).}\label{cv_flux}
\end{figure*}

\subsection{Reliability of our identifications}\label{false}
In order to test the reliability of our identifications, we replicate the test described in \S~\ref{selection} to estimate the probability of false matches for the stars included in our final catalog. We shift the positions of our stars by a few arcminutes and rematch the catalog to the RASS. This consistently returns fewer than five matches, indicating that the chance of a random association between a star in our catalog and a RASS source is under $1\%$. 

A more conservative estimate of the number of false identifications in our catalog comes from our comparison to the \citet{flesch04} catalog discussed above. There are roughly $300$ X-ray sources listed in that catalog for which we also provide identifications. For $\sim10\%$ of those our classifications disagree, and we take this to be an upper limit on the contamination rate of our catalog as a whole.   

\section{Using SDSS Spectroscopic Catalogs to Identify X-ray Emitting...}\label{wds_cvs}
\subsection{Cataclysmic Variables}
As evidenced from their small numbers in our SIMBAD and APO sample of RASS counterparts, CVs and WDs (discussed in \S~\ref{wds_details}) are both rarer types of stellar X-ray sources. In order to build up meaningful samples of these objects, we use the full SDSS spectroscopic catalogs for our correlations with the RASS.

The properties of the $177$ new CVs spectroscopically identified in SDSS have been described by \citet{paula1, paula2, paula3, paula4, paula5, paula6}; an additional $36$ previously known CVs have been recovered in the survey data. $45$ SDSS CVs have X-ray counterparts in the RASS; all but one have been published in the Szkody et al.\ papers (SDSS J100515.38$+$191107.9 will be included in the next catalog). Below we discuss the $(f_X/f_{opt})$ distributions for the X-ray emitting CVs and reproduce the \citet{joe_p1} log $(f_X/f_{opt})$ vs.\ H$\beta$ EqW plot for the $46$ for which we have spectra.

\subsubsection{Flux ratios and H$\beta$ equivalent widths}\label{cv_details}
We calculate log $(f_X/f_g)$ for the $45$ X-ray emitting SDSS CVs, for EI UMa, the only CV in our RASS/SIMBAD sample not included in the SDSS CV catalogs, and for SDSS J1714, the CV discovered in our APO spectroscopic survey. While this distribution is fairly symmetric about the median ($+0.38$) (left panel, Figure~\ref{cv_flux}), we find, as expected, an offset between the flux ratios for confirmed polars (median log $(f_X/f_g) = +0.99$) and for dwarf novae (median log $(f_X/f_g) = +0.13$) (right panel, Figure~\ref{cv_flux}).

\citet{joe_p1} described a correlation between $(f_X/f_V)$ and the EqW of the H$\beta$ emission line, a proxy for the CV's overall emission strength. In Figure~\ref{h_beta_vs} we plot these two quantities for our CV sample\footnote{The color-dependent difference between $g$ and $V$ is small: $g = V + 0.05$ for a low-redshift QSO with $B-V = 0.3$ \citep[][]{fukugita}.}. The H$\beta$ EqWs are reported in \citet{paula1, paula2, paula3, paula4, paula5, paula6} for all but two of these CVs (SDSS J083642.80$+$532838.1/SW UMa and SDSS J100515.38$+$191107.9); in those cases (and that of our new CV) we measured them using IRAF's {\it splot} routine.

The scatter around the $(f_X/f_g)$ predicted by the \citet{joe_p1} relationship is comparable to that found in other studies \citep[e.g.,][]{richman96}. However, there are clear departures from this relationship, with the polars appearing to form a distinct population from the dwarf novae and other CVs. Furthermore, the SDSS CVs tend to have higher $(f_X/f_{opt})$ ratio than those included in the earlier studies, where only a handful had log $(f_X/f_{opt}) \geq 0$ \citep{richman96}.

\subsubsection{Outliers}
\begin{figure}[h]
\epsscale{1.1}
\centerline{\plotone{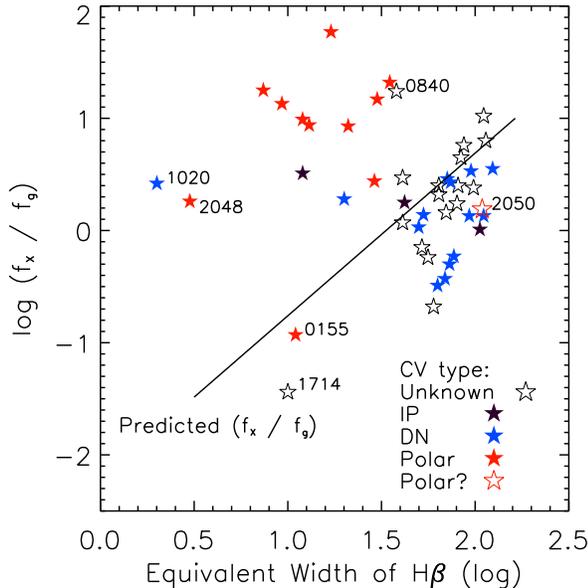}}
\caption{log $(f_X/f_g)$ vs.\ EqWs of the H$\beta$ line for the $46$ X-ray emitting SDSS CVs for which we have spectra. The predicted $(f_X/f_g)$ is from \citet{joe_p1}. The labeled objects are discussed in the text.}\label{h_beta_vs}
\end{figure}

SDSS J015543.40$+$002807.2 occupies an unusual position in Figure~\ref{h_beta_vs} for a polar because of its low X-ray-to-optical flux ratio (log $(f_X/f_g) = -0.93$). As noted by \citet{paula1}, SDSS J0155 was significantly brighter when observed by SDSS than in its Digitized Sky Survey image. A change in brightness of a few magnitudes--typical for polars entering a phase of stronger accretion--would be sufficient to explain the apparent inconsistency of this system's log $(f_X/f_g)$ relative to that of other polars. Follow-up observations by \citet[][]{woudt2004} and \citet{sch2005} confirmed the large changes in mass transfer rate in this system. 

The spectrum of SDSS J205017.84$-$053626.8 shows HeII emission, but initial follow-up observations of this CV could not confirm that this is a magnetic system \citep{paula2}. Its log $(f_X/f_g)$ of $+0.19$ is similar to that of other intermediate polars (IPs) in our sample, but SDSS J2050 has a period of $1.6$ hr, much shorter than is typical for IPs. \citet[][]{lee06} determined that this is an example of the most magnetized type of CV, a disk-less, stream-fed polar.

SDSS J102026.53$+$530433.1, the dwarf nova KS UMa, has an anomalously low H$\beta$ EqW relative to other DNe; this is because its SDSS spectrum was taken during an outburst and is dominated by the accretion disk. 

The spectrum of SDSS J204827.91$+$005008.9 is very unusual, with a hump apparent between $4000$ and $5000$ \AA\ and a typical M-dwarf continuum to the red of the hump \citep[see][]{paula5}. Follow-up observations by \citet{schmidt} led to the conclusion that this low-accretion magnetic binary is possibly a pre-polar. The X-ray luminosity derived for this CV is an order of magnitude greater than the $L_X$ for similar SDSS systems, where the coronae of the late-type companion stars are thought to be responsible for the X-ray emission \citep{paula2004}. \citet{schmidt} suggested that this weakens the case for SDSS J2048 as the $ROSAT$-detected X-ray source. Accordingly, on 2005 Aug 29 we obtained a DIS low-resolution spectrum for the most obvious potential X-ray counterpart in the RASS field, a bright nearby star ($g = 14.43$), which we then typed using the Hammer. This late G star has a 2MASS counterpart with $J = 11.01$; the resulting log $(f_X/f_J)$ of $-2.15$ is within the typical G star range (see Table~\ref{known_table}). While this suggests that the G star may in fact be the RASS counterpart, this is a crowded field; pointed X-ray observations are required to confirm that this star, rather than SDSS J2048, is the RASS-detected X-ray source.

Finally, when discovered, SDSS J084026.16$+$220446.6 did not appear to have strong HeII emission \citep{paula5}, but its log $(f_X/f_g)$ and H$\beta$ EqW are very similar to those of the known polar GG Leo. However, definitively classifying this system and many other SDSS CVs clearly requires more observational data than are provided by the survey. 

\subsection{White Dwarfs}\label{wds_details}
WDs cool as they age, and the most commonly observed WDs, DAs, have an effective temperature (T$_{eff}$) distribution that peaks at $\sim10^4$~K \citep[e.g.,][]{eisenstein06}. At that temperature, one would not expect a typical DA to be a significant producer of thermal X rays. However, \citet{shipman76} pointed out that the X rays produced in the hotter interior of a DA should travel through its thin hydrogen atmosphere unabsorbed. 

The consensus was that the RASS would include upwards of $5000$ X-ray emitting DAs \citep[e.g.,][]{barstow89}, but to date the number found in the RASS is far smaller. \citet{fleming96} searched $55,000$ RASS sources for the super-soft sources (HR1 $=-1$ and HR2 $\leq0$), found only $161$ DAs, and concluded that $90\%$ of DAs are opaque to X-ray emission. More recent RASS identification programs have added only a few dozen X-ray emitting WDs to this sample \citep[e.g.,][]{zickgraf03}. 

The SDSS DR4 produced a catalog of $9309$ spectroscopically confirmed WDs. $\sim8000$ of these are DAs  and $\sim 6000$ are new \citep{eisenstein06}. Correlating SDSS WD lists with the RASS does not add many new WDs to the sample of known X-ray emitters (see the Appendix for details about our correlations). However, among these new X-ray emitters are three that, if confirmed, may be the coolest X-ray emitting DAs detected (see \S~\ref{cool}), suggesting that these fainter SDSS X-ray emitting WDs, while small in number, may prove to be very interesting. Below we summarize our results before discussing the DAs' X-ray-to-optical flux ratio and temperature distributions.

\subsubsection{New SDSS X-ray emitting white dwarfs}
We list $17$ new confident DA WD/RASS source associations in Table~\ref{known_and_new_wds}; this number includes three post-DR4 WDs, described below, and four \citet{mccook99} DAs not previously identified as X-ray sources\footnote{Our matching returned an additional $11$ known X-ray emitting DAs; ten were identified by \cite{fleming96} or \cite{zickgraf03} or both; the other DA was reported as an X-ray emitter by \cite{chu04}.}. Table~\ref{known_and_new_wds} gives the relevant data for these DAs, including the \citet{mccook99} catalog name if available.

\begin{deluxetable*}{lclcccl}
\tablewidth{0pt}
\tabletypesize{\scriptsize}
\tablecaption{New X-ray emitting SDSS DA WDs\label{known_and_new_wds}}
\tablehead{
\colhead{}       &\colhead{T$_{eff}$} &\colhead{}       &\colhead{Offset} &\colhead{}             &\colhead{log} &\colhead{McCook \&}  \\
\colhead{SDSS J} &\colhead{(K)}         &\colhead{1RXS J} &\colhead{(\asec)}  &\colhead{Counts s$^{-1}$} &\colhead{$(f_X/f_g)$} & \colhead{Sion (1999)}
}
\startdata
075106.51$+$301726.4 & $33158\pm65$ & 075107.4$+$301719 & $13.7$ & $0.038\pm0.013$ & $-0.69$\tablenotemark{a} & \nodata \\ 
082346.14$+$245345.6 & $34521\pm47$ & 082345.4$+$245344 & $10.2$ & $0.039\pm0.014$ & $-0.81$ & \nodata \\
091215.43$+$011958.8 & $55691\pm1415$ & 091215.5$+$012003 & $4.8$ & $0.041\pm0.015$ & $-0.01$\tablenotemark{a} & WD 0909$+$015\\
093006.79$+$522803.3 & $36469\pm185$ & 093006.7$+$522752 & $11.4$ & $0.039\pm0.013$ & $-0.53$ & \nodata \\
093608.90$+$382200.5 & $37809\pm397$ & 093609.6$+$382237 & $37.4$ & $0.038\pm0.012$ & $+0.00$ & WD 0933$+$385\\
093921.83$+$264401.1\tablenotemark{b} & $59224\pm1253$ & 093922.8$+$264404 & $13.3$ & $0.044\pm0.013$ & $-0.17$ & \nodata \\
095245.58$+$020938.9 & $43114\pm443$ & 095244.6$+$020946 & $16.6$ & $0.025\pm0.010$ & $-0.69$ & WD 0950$+$023\\
100543.92$+$304744.9\tablenotemark{b} & $62299\pm1232$ & 100543.7$+$304809 & $24.5$ & $0.046\pm0.016$ & $-0.42$ & \nodata \\
101339.56$+$061529.5 & $14375\pm603$ & 101340.6$+$061556 & $30.6$ & $0.019\pm0.009$ & $+0.70$ & \nodata \\
113836.33$+$475509.8 & $50602\pm1169$ & 113836.9$+$475459 & $12.3$ & $0.046\pm0.015$ & $+0.04$ & \nodata \\
120432.67$+$581936.9 & $15725\pm545$ & 120428.0$+$581934 & $36.9$ & $0.022\pm0.010$ & $+0.63$ & \nodata \\
125938.08$+$603900.0 & $37640\pm225$ & 125936.6$+$603910 & $14.8$ & $0.039\pm0.012$ & $-0.30$ & \nodata \\
132629.58$+$571131.5 & $94382\pm5479$ & 132634.4$+$571052 & $55.7$ & $0.020\pm0.009$ & $+0.11$ & \nodata \\
143736.67$+$362213.4\tablenotemark{b} & \nodata & 143738.0$+$362237 & $36.4$ & $0.029\pm0.009$ & $-0.23$ & \nodata \\
145600.79$+$574150.7 & $31500\pm79$ & 145605.0$+$574132 & $38.6$ & $0.021\pm0.008$ & $-0.83$ & WD 1454$+$578\\
154448.25$+$455039.0 & $53109\pm1119$ & 154449.3$+$455035 & $11.5$ & $0.046\pm0.011$ & $-0.08$ & \nodata \\
163418.25$+$365827.1 & $10080\pm216$ & 163421.8$+$365859 & $53.0$ & $0.016\pm0.007$ & $+0.60$ & \nodata \\
\tableline
022618.60$-$083049.3 & $9692\pm153$ & 022615.8$-$083108 & $45.8$ & $0.033\pm0.011$ & $+0.94$ & WD 0223$-$087 \\
074859.06$+$215445.7 & $29717\pm911$ & 074858.9$+$215458 & $12.5$ & $0.026\pm0.010$ & $+0.75$ & \nodata \\
093307.65$+$400637.9 & $9037\pm57$ & 093305.6$+$400553   & $50.7$ & $0.035\pm0.011$ & $+0.46$ & \nodata \\
093456.44$+$023121.1 & $20844\pm1009$ & 093458.0$+$023151 & $37.9$ & $0.023\pm0.011$ & $+0.60$  & WD 0932$+$027 \\ 
103237.35$+$512857.9 & $20983\pm585$ & 103236.8$+$512809 & $48.7$ & $0.016\pm0.007$ & $+0.35$ & \nodata \\
112740.90$-$024638.7 & $20784\pm206$ & 112741.1$-$024614 & $24.9$ & $0.017\pm0.007$ & $-0.45$ & \nodata \\
152349.89$+$041434.6 & $23852\pm542$ & 152351.0$+$041356 & $41.6$ & $0.059\pm0.020$ & $+0.49$ & WD 1521$+$044 \\
165645.49$+$182437.7 & $12460\pm638$ & 165643.7$+$182441 & $25.8$ & $0.015\pm0.006$ & $+0.66$ & \nodata \\
232658.89$-$002339.9 & $8081\pm83$ & 232701.2$-$002337   & $34.6$ & $0.018\pm0.009$ & $+0.37$ &  \nodata \\
232659.23$-$002348.0 & $10622\pm47$ & 232701.2$-$002337  & $31.3$ & $0.018\pm0.009$ & $-0.37$ & \nodata
\enddata
\tablecomments{Above the line are highly secure RASS counterpart identifications; below it are tentative identifications requiring further observations. SDSS J0936 is listed as a DQ WD by \citet{mccook99}. However, the SDSS spectrum does not show the carbon absorption lines typical of DQs, and it seems that this is indeed a DA WD.}
\tablenotetext{a}{$g$ photometry is flagged in SDSS.} 
\tablenotetext{b}{Post-DR4 WD.}
\end{deluxetable*}

Three of these DAs are not included in the \citet{eisenstein06} DR4 catalog: SDSS J093921.83$+$264401.1, J100543.92$+$304744.9, and J143736.67$+$362213.4. We have therefore fitted the SDSS spectra using the WD models described in \citet{kleinman04} to obtain estimates for their T$_{eff}$ and log g. Because of calibration problems with the SDSS spectrum, this fitting fails for SDSS J1437, and we identify it as a DA based on visual inspection of the spectrum. For SDSS J0939, the fits yield that the star is a DA$0.9$ with log g $=7.9$ and T$_{eff} \sim 59,000$ K; SDSS J1005 is a DA$0.8$ with log g $=7.6$ and T$_{eff} \sim 62,000$ K (see Figure~\ref{DA_fit1}). 

We also identify ten DAs as less certain RASS counterparts and list them in Table~\ref{known_and_new_wds} for reference. Follow-up optical and/or X-ray observations are required to confirm these associations.

We further identify several WDs of other types as RASS counterparts. Among these are three DAs with M star companions, two DB WDs, which have helium-rich atmospheres, and a PG 1159 star, a pre-WD star showing strong oxygen and carbon lines. We also identify two other non-DA WDs from our post-DR4 spectra as likely RASS sources. We list these objects in Table~\ref{other_wds} and discuss follow-up {\it Chandra} observations of the two candidate X-ray emitting DBs in \S~\ref{dbs}.

\subsubsection{log $(f_X/f_g)$ and temperature distributions}\label{cool}
\begin{figure}
\epsscale{1.1}
\centerline{\plotone{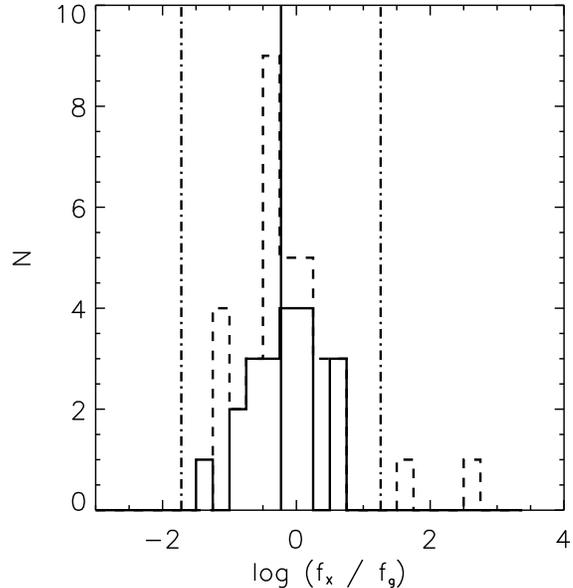}}
\caption{log $(f_X/f_g)$ distributions for X-ray emitting DA WDs. The dashed histogram is for the $36$ \citet{fleming96} DAs in the SDSS footprint; the solid one is for the sample identified here (eight recovered \citet{fleming96} DAs are not included in the latter histogram). The solid vertical line is the mean flux ratio, and the dot-dashed vertical lines bracket the $2 \sigma$ ranges, for the \citet{fleming96} sample.}
\label{flux}
\end{figure}

In Figure~\ref{flux} we plot the log $(f_X/f_g)$ distribution for the $36$ \citet{fleming96} X-ray emitting DA WDs within the SDSS footprint\footnote{This includes the six WDs from Table~\ref{wd_sim} listed in the \cite{fleming96} catalog. Of the eight WDs in that table, only the last, WD 1725$+$586, has an SDSS spectrum. It is therefore the only X-ray emitting WD identified in our earlier work included in the rest of this analysis.}. The median log $(f_X/f_g) = -0.23$, with the $2\sigma$ range being from $-1.72$ to $+1.26$. The $20$ WDs not cataloged in \citet{fleming96} that we have confidently identified as X-ray emitters have flux ratios that fall within this range. Although not included in this Figure, the $10$ additional less certain X-ray emitting DAs listed in Table~\ref{known_and_new_wds} also have flux ratio values within this range.

\begin{figure}
\epsscale{1.1}
\centerline{\plotone{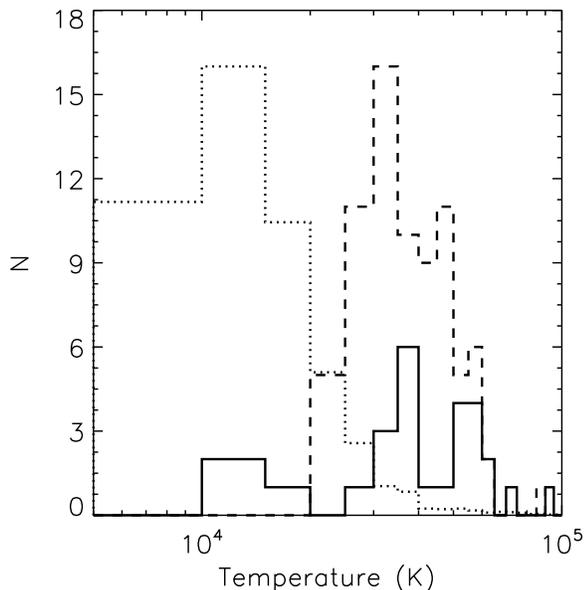}}
\caption{Temperature distributions for X-ray emitting and ``ordinary'' DA WDs. The dashed histogram is the \citet{fleming96} sample of X-ray emitters; the solid one is the sample identified from SDSS/RASS matching. The dotted histogram is the normalized distribution for $7578$ \citet{eisenstein06} DAs.}
\label{temps}
\end{figure}

In Figure~\ref{temps} we compare the temperature distribution for our X-ray emitting DAs to that of the $76$ \citet{fleming96} DAs for which \citet{marsh97a} obtained temperatures. We include $15$ previously known and $12$ new DAs for which the association with a RASS source is most secure\footnote{One new DA, SDSS J1437, lacks a T$_{eff}$, as mentioned above.}. Only three \citet{fleming96} DAs in our sample have \citet{marsh97a} temperatures and are plotted twice: WD 1026$+$453, 1429$+$373, and 1725$+$5863. For these DAs, the difference between the \citet{eisenstein06} and \citet{marsh97a} T$_{eff}$ are rather large ($2700$, $1900$, and $8900$ K), with the \citet{eisenstein06} T$_{eff}$ being greater in each case. Discrepancies such as these have been noticed in comparisons between SDSS temperatures and those in the literature, particularly for the hotter objects, with the differences reaching $\sim10\%$ at $50,000$~K (D.~Eisenstein 2006, private communication). Nevertheless, the temperature distributions for our sample of X-ray emitting DAs (median T$_{eff} = 43,100$ K) and that of \citet{fleming96} are very similar.

We also plot the T$_{eff}$ distribution for $7578$ \citet{eisenstein06} DAs, normalized to the maximum of the \citet{fleming96} distribution. The separation between X-ray emitters and ``ordinary'' WDs is clear, with the \citet{fleming96} DAs having a median T$_{eff} \sim39,000$~K, compared to $14,000$~K for the \citet{eisenstein06} DAs. Interestingly, $1251$ of the \citet{eisenstein06} DAs are hotter than $22,153$~K, the temperature of the coolest \citet{fleming96} DA. Even if, as suggested by \citet{fleming96}, $90\%$ of DAs are opaque to X rays, we would expect $\sim125$ of these to be X-ray sources. If we include the uncertain WD/RASS associations, we have to date found roughly half that number. One possible explanation for this discrepancy is that the use of the \citet{fleming96} super-softness criterion is too restrictive at fainter X-ray fluxes, where the hardness ratios have larger errors. Alternatively, these sources may be too faint to be detected in the RASS. We plan to explore this question further.

Three of our new X-ray emitting DAs are cooler than the coolest \citet{fleming96} DA and below the minimum temperature theoretically  needed for X-ray emission from a DA. The three DAs, SDSS J101339.56$+$061529.5, J120432.67$+$581936.9, and J163418.25$+$365827.1, are faint ($g > 19$ mag), and the quoted T$_{eff}$ should be considered with caution. That said, because of the strength of the Balmer absorption lines, model fits to lower temperature SDSS WD spectra are very reliable  (D.~Eisenstein 2006, private communication). Follow-up optical and especially X-ray observations, given the low X-ray detection likelihoods of the RASS sources, are called for.

One of these DAs, SDSS J1634, was observed with {\it XMM} on 2008 Mar 11 (ObsId 0503550701). The X-ray background for these observations was very high; the on-source integration time was very short ($2200$ s rather than the $8000$ s requested) and only MOS data are available. As the MOS is not very sensitive at soft energies, the non-detection of the WD in these data is neither surprising nor conclusive. 

\begin{deluxetable*}{lclccccl}
\tablewidth{0pt}
\tabletypesize{\scriptsize}
\tablecaption{X-ray emitting SDSS non-DA WDs\label{other_wds}}
\tablehead{
\colhead{}     & \colhead{T$_{eff}$} & \colhead{}         & \colhead{Offset}        &  \colhead{}               & \colhead{log}        &\colhead{}  & \colhead{McCook \&} \\
\colhead{SDSS J} & \colhead{(K)}       & \colhead{1RXS J} &\colhead{(\asec)} & \colhead{Counts s$^{-1}$} & \colhead{$(f_X/f_g)$} & \colhead{Type}   & \colhead{Sion (1999)} 
}
\startdata
093122.86$+$362209.4 & $20496\pm1050$ & 093119.2$+$362223 & $46.4$ & $0.017\pm0.008$ & $+0.60$ & DB  & \nodata\\
153225.49$+$472700.9 & $17567\pm447$  & 153221.7$+$472648 & $40.4$ & $0.012\pm0.005$ & $+0.20$ & DB  & \nodata \\
114635.23$+$001233.5 & $100000\pm794$\tablenotemark{a} & 114634.6$+$001238 & $10.7$ & $0.091\pm0.020$ & $-0.72$ & PG 1159  & \nodata\\
114312.57$+$000926.6 & $17596\pm607$  & 114313.8$+$000955 & $34.2$ & $0.038\pm0.017$ & $+0.22$ & DA+dM6e & \nodata\\
132725.72$+$505711.2\tablenotemark{b} & $10848\pm112$\tablenotemark{a} & 132725.7$+$505713 & $2.0$ & $0.047\pm0.012$ & $+0.04$ & DA+dM & WD 1325$+$512 \\
134100.03$+$602610.4 & $46306\pm0$    & 134101.0$+$602556 & $15.7$ & $0.034\pm0.009$ & $-0.42$ &  DA+dM3e & \nodata\\
141740.23$+$130148.7\tablenotemark{b} & $36000\pm51$ & 141741.2$+$130206 & $22.5$ & $0.032\pm0.012$ & $-1.11$ & DA+dM & WD 1415$+$132 \\
223051.15$+$125706.8 & $21125\pm880$  & 223051.6$+$125713 & $9.4$ & $0.023\pm0.010$ & $+0.55$ & DA+dM4e & \nodata
\enddata
\tablenotetext{a}{Fit for T$_{eff}$ is unreliable.}
\tablenotetext{b}{Post-DR4 WD.}
\end{deluxetable*}

\subsubsection{Observations of candidate cool X-ray emitting DBs}\label{dbs}
The first two WDs listed in Table~\ref{other_wds} are DB white dwarfs newly identified by \citet{eisenstein06}. These DBs are cooler than one would expect for non-DA X-ray emitters \citep[cf.\ discussion in][]{odwyer03}; however, both are at relatively large offsets from the corresponding RASS sources. Furthermore, these RASS sources have low detection likelihoods; visual inspection of the RASS images suggests that both sources were marginally detected. $5$ ksec {\it Chandra} snapshots of SDSS J0931 on 2008 Jan 16 (\dataset[ADS/Sa.CXO#Obs/8916]{ObsId 8916}) and J1532 on 2008 Jun 01(\dataset[ADS/Sa.CXO#Obs/8917]{ObsId 8917}) were therefore taken to locate the RASS source and test whether the DBs are in fact X-ray sources. Sources were identified using the Ciao routine {\it dmextract}.

There is no {\it Chandra} source at the position of the WD in either observation, nor do we appear to have detected either of the RASS sources. We discuss each observation below; we conclude that the RASS sources were either spurious or transient.

\begin{itemize}
\item {\it SDSS J0931} The closest source to the RASS position is a weekly detected ($<3\sigma$) optically faint ($g=22.1$ mag) galaxy\footnote{Star/galaxy separation at these magnitudes is unreliable \citep{scranton02}}. Taking into account the photometric errors, SDSS J$093119.70+362202.4$ has $u-g\ \lapprox\ 0.6$, giving it the UV-excess typically seen in SDSS QSOs \citep[cf.][]{richards02}; its log $(f_X/f_g)$ of $1.0$ is consistent with this being an AGN or QSO. While it is offset by $22\asec$ from the RASS position, it seems unlikely that this is the RASS source, however, as its {\it Chandra} count rate is substantially less than that predicted from the RASS source count rate. 

We note the serendipitous detection of the counterpart to the RASS source 1RXS J$093210.1+363010$. SDSS J$093209.60+363002.5$ is a relatively bright ($g = 17.8$ mag\footnote{This is the Petrosian magnitude.}) z $=0.15$ early type galaxy lacking both the photometric UV-excess and the strong spectroscopic emission lines typically seen in AGN. The likely counterpart to a FIRST source, this is plausibly an X-ray emitting BL Lac, as suggested by \citet{rich08}, although it also has some similarities to e.g., CXOU J031238.9$-$765134, an X-ray bright optically normal galaxy \citep[XBONG;][]{comastri02}.

\item  {\it SDSS J1532} The closest {\it Chandra}-detected source corresponds to a faint SDSS object ($g = 21.1$ mag). Within the photometric errors, SDSS J$153244.22+472405.2$ has $u-g < 0.6$; its log $(f_X/f_g)$ of $0.8$ is consistent with this being an AGN or QSO. However, this object is over $4\amin$ from the RASS position and its {\it Chandra} count rate is significantly less than that predicted based on the RASS source count rate. 
\end{itemize}

\section{Conclusion}\label{concl}
We have used photometric and spectroscopic data from SDSS to identify $709$ stellar X-ray emitters detected in the $\ROSAT$ All-Sky Survey. The majority of these are optically bright stars with coronal X-ray emission, and were therefore not identified based on routine SDSS spectroscopy, which typically does not target $\lapprox\ 15$ mag objects. Instead, we used SDSS imaging photometry, correlations with 2MASS and other catalogs, and spectroscopy from the APO $3.5$-m telescope to identify these stellar X-ray counterparts. Our sample of $707$ X-ray emitting F, G, K, and M stars characterized in this fashion is one of the largest X-ray selected samples of such stars. Our catalog also includes a known WD, WD 1725$+$586, not previously identified as an X-ray source, as well as a new X-ray emitting CV, SDSS J171456.78$+$585128.3, identified on the basis of follow-up APO observations. We estimate that our contamination rate by false matches between a RASS source and a star is $\lapprox\ 10\%$. 

In order to expand the sample of X-ray emitting WDs and CVs, we also matched the SDSS spectroscopic catalogs of these objects \citep[i.e.,][]{eisenstein06, paula6} to the RASS. We identified $44$ of the SDSS CVs published to date as RASS sources; roughly half are SDSS discoveries, with the others being previously known systems. We add a new X-ray emitting CV to this list; SDSS J1005$+$1911 is among the CVs spectroscopically identified since the last SDSS catalog \citep{paula6} and is a SDSS discovery.

Correlating the SDSS DR4 WD catalog of \citet{eisenstein06} with the RASS does not yield very many new entries in the catalog of known X-ray emitting WDs. We identify $25$ DA WDs included in \citet{eisenstein06} as RASS sources; $14$ of these are new identifications (e.g., not included in the \citet{fleming96}, \citet{zickgraf03}, and/or \citet{chu04} catalogs of X-ray emitting WDs). Examination of post-DR4 spectroscopy allows us to identify three additional SDSS-discovered DAs as RASS counterparts, but the total number of X-ray emitting WDs in SDSS remains small.

This dearth of new X-ray emitting WDs may be a result of the relative sensitivities of the two surveys. Stellar counterparts to RASS sources tend to be very bright optically, and can be identified without requiring the depth of SDSS--while the fainter WDs discovered by SDSS are unlikely to be bright enough in the X-ray regime to have RASS detections (e.g., because they are too distant). However, among the new RASS WDs are three that, if confirmed, may be the coolest X-ray emitting DAs detected to date, suggesting that these fainter SDSS X-ray emitting WDs, while small in number, may prove to be very interesting. In these cases, as in the cases of other, less certain WD/RASS associations we propose, follow-up X-ray and/or optical observations are necessary.

\begin{deluxetable*}{lclcccc}
\tablewidth{0pt}
\tabletypesize{\scriptsize}
\tablecaption{RASS/SIMBAD Class II, III, IV, and A stars\label{other_L}}
\tablehead{
\colhead{ }      &\colhead{ }               &\colhead{ }    & \colhead{Offset}  & \colhead{$V$}   & \colhead{log }        & \colhead{}     \\
\colhead{1RXS J} &\colhead{Counts s$^{-1}$} &\colhead{Name} & \colhead{(\asec)} & \colhead{(mag)} & \colhead{$(f_X/f_V)$} & \colhead{Type}
}
\startdata
012041.7$+$154141 & $0.162\pm0.027$ & HD 8110    & $9.6$ & $7.28$ & $-3.51$ & G8IV \\
015332.2$+$124047 & $0.028\pm0.010$ & HD 11520   & $34.6$ & $7.06$ & $-4.36$ & F6III \\
082929.4$+$453210 & $0.040\pm0.011$ & BD+46 1404 & $3.0$ & $10.14$ & $-2.97$ & K0III \\
113615.1$+$662047 & $0.018\pm0.007$ & CCDM J11362+6621AB  & $5.29$ & $7.91$ & $-4.20$ & F5IV; $**$ \\
141050.6$+$623121 & $0.018\pm0.006$ & HD 124370  & $17.0$ & $8.29$ & $-4.06$ & F2IV \\
164532.0$+$422651 & $0.044\pm0.010$ & HD 151445  & $1.6$ & $8.15$ & $-3.73$ & F5IV \\
165947.2$+$375045 & $0.028\pm0.008$ & HD 153777  & $35.2$ & $8.47$ & $-3.79$ & F2IV \\
171558.3$+$270804 & $0.026\pm0.008$ & HD 156362  & $8.1$ & $6.58$ & $-4.58$ & K2III \\
171701.2$+$264252 & $0.032\pm0.009$ & HD 156536  & $8.3$ & $7.51$ & $-4.13$ & F3IV \\
\tableline
013552.3$+$000559 & $0.025\pm0.010$ & HD 9806  & $17.7$ & $9.39$ & $-3.48$ & A2 \\
032728.1$-$065748 & $0.063\pm0.015$ & HD 21468  & $5.2$ & $9.07$ & $-3.20$ & A2 \\
093642.4$+$565820 & $0.175\pm0.019$ & HD 82861  & $8.4$ & $7.06$ & $-3.56$ & Am \\
130412.8$+$031257 & $0.027\pm0.012$ & HD 113498  & $36.3$ & $9.55$ & $-3.37$ & A2 \\
132218.6$+$055348 & $0.039\pm0.016$ & V$*$ BC Vir  & $53.3$ & $11.9$ & $-2.28$ & A; RR$*$ \\
140251.4$+$614051 & $0.012\pm0.005$ & HD 122944  & $32.1$ & $7.30$ & $-4.64$ & A2 \\
144952.6$+$035517 & $0.020\pm0.011$ & HD 130770  & $14.5$ & $8.29$ & $-4.01$ & A3 \\
155108.2$+$525439 & $0.118\pm0.014$ & HD 142282  & $21.9$ & $6.5$ & $-3.96$ & A0; $*$i$*$ \\
163147.7$+$453527 & $0.041\pm0.009$ & BD+45 2422B  & $14.0$ & $8.89$ & $-3.46$ & A0; $*$i$*$ \\
165344.9$+$630608 & $0.017\pm0.003$ & HD 153204  & $18.4$ & $8.55$ & $-3.98$ & A2 \\
171337.2$+$304151 & $0.011\pm0.005$ & V$*$ TU Her  & $50.8$ & $11.2$ & $-3.09$ & A5; Al$*$ 
\enddata
\tablecomments{``$**$'' or ``$*$i$*$'' indicates a double or multiple star system, ``RR*'' a RR Lyrae star, and ``Al$*$'' an Algol-type binary.}
\end{deluxetable*}

\acknowledgements
We thank the referee for comments that improved the paper. We are grateful to the APO time allocation committee at the University of Washington for providing us with the nights needed to complete our spectroscopic program. We thank the telescope operators at the observatory for their assistance with our observations. We also thank Michelle Cash, Nick Cowan, Lee Mannikko, Anil Seth, and Kristine Washburn for using some of their APO observing time to obtain data for this program. Finally, we thank Beth Willman for sharing her matching code. M.\ Ag\"ueros is supported by an NSF Astronomy and Astrophysics Postdoctoral Fellowship under award AST-0602099; much of the work described here was made possible by a NASA Harriett G.\ Jenkins Predoctoral Fellowship. Further NASA support was provided to K.\ Covey through the Spitzer Space Telescope Fellowship Program, through a contract issued by the Jet Propulsion Laboratory, California Institute of Technology under a contract with NASA. We gratefully acknowledge {\it Chandra} grants GO6-7058X and GO8-9012X for support of portions of our program. 

This research has made use of data obtained from the Chandra Data Archive and software provided by the Chandra X-ray Center (CXC) in the application packages CIAO, ChIPS, and Sherpa. Some of this work is based on observations obtained with XMM-Newton, an ESA science mission with instruments and contributions directly funded by ESA Member States and NASA.

Funding for the SDSS and SDSS-II has been provided by the Alfred P. Sloan Foundation, the Participating Institutions, the National Science Foundation, the U.S. Department of Energy, the National Aeronautics and Space Administration, the Japanese Monbukagakusho, the Max Planck Society, and the Higher Education Funding Council for England. The SDSS Web Site is http://www.sdss.org/.

The SDSS is managed by the Astrophysical Research Consortium for the Participating Institutions. The Participating Institutions are the American Museum of Natural History, Astrophysical Institute Potsdam, University of Basel, University of Cambridge, Case Western Reserve University, University of Chicago, Drexel University, Fermilab, the Institute for Advanced Study, the Japan Participation Group, Johns Hopkins University, the Joint Institute for Nuclear Astrophysics, the Kavli Institute for Particle Astrophysics and Cosmology, the Korean Scientist Group, the Chinese Academy of Sciences (LAMOST), Los Alamos National Laboratory, the Max-Planck-Institute for Astronomy (MPIA), the Max-Planck-Institute for Astrophysics (MPA), New Mexico State University, Ohio State University, University of Pittsburgh, University of Portsmouth, Princeton University, the United States Naval Observatory, and the University of Washington.

This publication makes use of data products from the Two Micron All Sky Survey, which is a joint project of the University of Massachusetts and the Infrared Processing and Analysis Center/California Institute of Technology, funded by the National Aeronautics and Space Administration and the National Science Foundation. 

\begin{center}
    {\bf APPENDIX}
  \end{center}
\section*{Non-Main Sequence And A Stars}\label{ap_1}
Table~\ref{other_L} gives the basic properties of the nine SIMBAD stars within $1$\amin\ of a RASS source cataloged as belonging to luminosity classes II, III, and IV that have not previously been associated with an X-ray source. We recovered an additional $14$ previously identified X-ray emitting members of these luminosity classes \citep[from][]{fleming89,haisch94,fava95,hempelmann96,appenzeller98,hunsch1998a,hunsch1998b}.

Table~\ref{other_L} also includes these data for the $11$ unpublished RASS/SIMBAD A stars. A stars are not generally predicted to be X-ray emitters; an unseen, young, active later-type companion is usually thought to be the source for the observed X rays \citep[c.f.\ discussion in][]{gudel04}. While we recover seven A stars that are previously known X-ray sources \citep[from][]{stocke91,simon95,shaw96,appenzeller98,hunsch1998b,schroder07}, it is unlikely, therefore, that all the other matched A stars are X-ray sources. However, the available SDSS data do not allow us to propose other counterparts to these X-ray sources. Follow-up X-ray and/or optical observations are required to test whether they are in fact the RASS sources. 

\begin{deluxetable*}{llclcccc}
\tablewidth{0pt}
\tabletypesize{\scriptsize}
\tablecaption{SIMBAD stars with APO spectra\label{simbad_apo}}
\tablehead{
\colhead{      } & \colhead{      } & \colhead{Offset}  & \colhead{SIMBAD} & \colhead{Offset}  & \colhead{Published} & \colhead{First APO} & \colhead{APO} \\
\colhead{1RXS J} & \colhead{SDSS J} & \colhead{(\asec)} & \colhead{name  } & \colhead{(\asec)} & \colhead{Type} & \colhead{Observation} & \colhead{Type}
}
\startdata
002552.2$-$095756 & 002550.97$-$095740.0 & $24.4$ & NLTT 1354 & $0.2$ & M3 & 2003 Sep 03 & M4 \\
010257.2$-$095145 & 010257.42$-$095140.5 & $8.8$  & HD 6204 & $0.1$ & G0 & 2003 Sep 03 & G3 \\
012724.2$-$093359 & 012723.80$-$093409.1 & $11.3$ & FBS 0124$-$098  & $0.1$ & M3...  & 2003 Nov 08 & M3 \\
020000.4$+$003216 & 015959.78$+$003219.3 & $9.8$  & [BHR2005] 284 & $0.4$ & M3.5V & 2003 Nov 16 & M3 \\
020144.5$-$101743 & 020143.87$-$101728.9 & $16.8$ & NLTT 6782 & $0.3$ & M3... & 2003 Nov 08 & M4 \\
020221.0$-$010715 & 020223.99$-$010708.3 & $45.4$ & BD$-$01 277 & $0.4$ & K0 & 2005 Nov 20 & K3  \\
021042.8$-$093817 & 021044.57$-$093826.0 & $27.7$ & NLTT 7231 & $0.0$ & M4 & 2003 Nov 08 & M3 \\
021559.9$-$092913 & 021558.94$-$092912.4 & $14.1$ & EUVE J0215$-$09.5\tablenotemark{a} & $2.5$ & Me & 2004 Oct 04 & M3 \\
022839.4$-$093550 & 022839.68$-$093545.5 & $6.5$  & HD 15424  & $0.3$ & F8 & 2005 Jan 05 & F2 \\
022942.6$-$003026 & 022941.79$-$003029.2 & $12.5$ & BD$-$01 341a & $1.0$ & G & 2003 Sep 03 & G8 \\
025150.0$-$081338 & 025148.21$-$081342.3 & $26.8$ & FBS 0249$-$084 & $0.3$ & M5... & 2004 Oct 04 & M4 \\
031114.2$+$010655 & 031115.49$+$010630.8 & $31.4$ & [BHR2005] 344 & $0.2$ & M5.5V & 2005 Nov 20 & M6 \\
101627.8$-$005127 & 101626.93$-$005139.4 & $17.6$ & RX J1016.4$-$0051\tablenotemark{b} & $2.4$ & M0e & 2004 Dec 03 & M1 \\
103546.1$+$021559 & 103546.90$+$021558.3 & $12.1$ & RX 1035.7$+$0215\tablenotemark{b} & $1.1$ & M2e & 2004 Dec 03 & M3 \\
104743.9$+$652750 & 104743.71$+$652736.0 & $14.1$ & HD 93270 & $0.3$ & F8 & 2005 Jan 29 & F7 \\
115749.7$+$663406 & 115749.80$+$663344.1 & $21.9$ & GSC 04160$-$00286 & $0.3$ & M & 2004 Dec 03 & M4 \\
142049.4$+$604930 & 142049.51$+$604934.9 & $4.5$  & RX J1420.8$+$6049\tablenotemark{c} & $3.9$ & M & 2004 Apr 30 & M3 \\
144243.9$-$003955 & 144244.40$-$003955.4 & $7.6$  & BD$-$00 2862\tablenotemark{d} & $0.3$ & G5V+K1V & 2005 Mar 01 & G2 \\
152346.8$-$004434 & 152346.11$-$004424.8 & $14.2$ & TYC $5003-309-1$\tablenotemark{e} & $1.7$ & Ke & 2005 Mar 30 & K1e \\
153826.3$+$523407 & 153826.38$+$523418.2\tablenotemark{f} & $11.3$ & HD 234250 & $1.7$ & K2 & 2005 Apr 26 & K3 \\
153912.0$+$554357 & 153912.04$+$554355.2 & $1.6$  & HD 238488 & $0.4$ & F8 & 2005 Apr 26 & F8 \\
165110.5$+$355500 & 165109.95$+$355507.2 & $9.5$  & NLTT 43695 & $0.2$ & M4 & 2005 May 31 & M5 \\
165445.3$+$423237 & 165445.04$+$423227.5 & $9.9$  & RX J1654.7$+$4232\tablenotemark{c} & $0.1$ & M & 2005 May 31 & M3 \\
170352.9$+$321147 & 170352.87$+$321146.2 & $0.9$  & TYC $2594-1478-1$\tablenotemark{a} & $0.6$ & M2.5 & 2005 Jul 19 & M3 \\
172142.4$+$620036 & 172142.60$+$620032.8 & $4.0$  & RX J1721.7$+$6200\tablenotemark{g} & $4.0$ & K4 & 2005 Sep 14 & K2 \\
172439.2$+$644054 & 172438.98$+$644052.5 & $2.0$  & TYC $4206-1161-1$\tablenotemark{g} & $0.1$ & G0 & 2005 Jul 20 & G7 \\
230240.3$+$003453 & 230241.41$+$003450.2 & $17.0$ & [ZEH2003] RX J2302.6$+$0034 3\tablenotemark{c} & $1.3$ & K & 2005 Nov 09 & K7e \\
\tableline
074553.1$+$355818 & 074555.71$+$355836.0 & $36.5$ & HD 62547  & $2.4$ & K0  & 2005 Jan 05 & G7 \\
081546.8$+$460123 & 081548.99$+$460119.5 & $23.2$ & [ZEH2003] RX J0815.7$+$4601 1\tablenotemark{c} & $3.2$ & F/G & 2005 Jan 29 & M2 \\
091609.6$+$015320 & 091610.17$+$015308.8 & $14.1$ & [ZEH2003] RX J0916.1$+$0153 3\tablenotemark{c} & $3.4$ & K & 2005 Jan 29 & M4 \\
092023.1$+$021351 & 092023.40$+$021340.5 & $11.4$ & [ZEH2003] RX J0920.3$+$0213 1\tablenotemark{c} & $0.5$ & F/G & 2005 Jan 29 & K7e \\
101716.7$+$051145 & 101716.76$+$051149.8 & $4.4$  & [ZEH2003] RX J1017.2$+$0511 2\tablenotemark{c} & $1.0$ & F/G & 2005 Jan 29 & G5 \\
102547.0$+$033143 & 102545.64$+$033141.0 & $21.0$ & [ZEH2003] RX J1025.7$+$0331 2\tablenotemark{c} & $0.7$ & F/G & 2004 Mar 28 & G2 \\
105706.3$-$010109 & 105707.70$-$010115.5 & $21.9$ & RX J1057.1$-$0101\tablenotemark{b,h} & $3.4$ & K4V & 2004 Dec 03 & G5 \\
110046.1$+$052307 & 110046.08$+$052304.6 & $2.4$  & [ZEH2003] RX J1100.7$+$0523 1\tablenotemark{c} & $1.5$ & K & 2004 Mar 28 & M3 \\
110804.1$+$052216 & 110803.94$+$052214.3 & $2.9$  & RX J1108.0$+$0522\tablenotemark{c} & $2.9$ & F/G & 2004 Mar 28 & M0 \\
143139.0$+$004732 & 143138.92$+$004731.4 & $1.5$  & RX J1431.6$+$0047\tablenotemark{c} & $2.7$ & K & 2005 Mar 01 & M0 \\
153709.0$+$531928 & 153709.43$+$531920.3 & $9.0$  & StKM 1$-$1262 & $2.7$ & K5 & 2005 Feb 26 & M0 \\
160602.0$+$501124 & 160602.24$+$501113.0 & $11.7$ & V$*$ V842 Her & $0.3$ & F9V & 2005 Apr 26 & G2 \\
161022.6$+$450934 & 161022.24$+$450934.9 & $3.8$  & RX J1610.3$+$4509\tablenotemark{c} & $4.9$ & K & 2004 May 22 & M0 \\
171928.8$+$652227 & 171928.68$+$652229.7 & $2.4$  & RX J1719.4$+$6522\tablenotemark{g} & $2.4$ & K0 & 2005 May 31 &  G7 
\enddata
\tablenotetext{a}{X-ray source identified by \citet{schwope}.}
\tablenotetext{b}{... by \citet{appenzeller98}.}
\tablenotetext{c}{... by \citet{zickgraf03}.}
\tablenotetext{d}{... by \citet{metanomski98}.}
\tablenotetext{e}{... by \citet{fischer98}.}
\tablenotetext{f}{One of two possible counterparts to this source (see Table~\ref{multiple_m}). This star was not included in the flux ration calculations discussed in \S~\ref{ratio_calc}.}
\tablenotetext{g}{... by \citet {gioia03}.}
\tablenotetext{h}{APO target is not the same star.}
\tablecomments{The first offset is between the RASS and SDSS positions; the second is between the SDSS and SIMBAD positions. Stars above the line are those for which choosing the APO or the SIMBAD spectral class does not impact $(f_X/f_{opt})$ calculations.}
\end{deluxetable*}

In addition, we identified two candidate A star counterparts to RASS sources in our APO spectroscopic campaign. In both cases, however, the A star seems unlikely to be the RASS source. TYC $3071-1446-1$, listed in SIMBAD without a spectral type, appears to be in front of a cluster of galaxies that includes 2MASX J$16571256+3920176$, identified as the RASS source RXS J$165711.6+392028$. 

The SDSS spectrum of SDSS J095220.07$+$531455.6, $33.5$\asec\ away from RXS J$095223.8+531455$, is that of a z $= 0.12$ QSO. While the candidate stellar counterpart is closer ($19.2$\asec), the RASS hardness ratios (HR1 $=-0.05\pm0.34$; HR2 $=-0.15\pm0.46$) are consistent with those commonly found for AGN \citep[e.g.,][]{voges99}, and it therefore seems most likely that the QSO is in fact the RASS source.

\section*{Comparison of SIMBAD and APO stellar classifications}\label{ap_1b}
In Table~\ref{simbad_apo} we list all the SIMBAD-cataloged candidate RASS counterparts with a spectral type for which we obtained an APO spectrum. For $27$ stars the spectral type derived from our spectrum either does not differ significantly from the SIMBAD type or is within the same spectral class. In these cases choosing one type or the other has no impact on the flux ratio calculations, since e.g., all F stars are binned together. In $15$ of the $20$ cases where subtypes are available in SIMBAD, the DIS-derived and SIMBAD spectral classes agree to within two subtypes. (The DIS spectra allow us to assign a subtype to seven other stars for which none was available.) 

For another $14$ stars, the choice of stellar type does impact the flux ratio calculations. In nine cases the star's type was derived by \citet{zickgraf03} from low-resolution spectra; we consider our types, obtained from higher resolution spectra, to be more reliable. In the five other cases, a stellar subtype is listed in SIMBAD; our DIS-derived spectral type agrees to within three subtypes in four cases--only slightly worse than what we consider our typical uncertainty when typing with the Hammer. The fourth case is that of the counterpart to 1RXS J105706.3$-$010109, identified by \citet{appenzeller98} as the K4V star RX J1057.1$-$0101. Our APO target was a star offset by $\sim5\asec$ from this one, explaining the discrepancy in spectral types.

$22$ of these stars are previously identified X-ray sources; three additional stars are listed in SIMBAD as RASS counterparts, but the origin of these associations is unclear. 

\section*{SDSS WD/RASS Identifications}\label{ap_3}
\section*{Building the sample of X-ray emitting DAs}
Table~\ref{known_and_new_wds} lists our most confident SDSS DA/RASS associations. We discuss how we obtained these and our less certain identifications. Ten of these less certain cases are also listed in Table~\ref{known_and_new_wds}; follow-up optical and/or X-ray observations are required to confirm these identifications.\\

\noindent {\it DAs discovered before SDSS}\\
\indent In correlating the \citet{eisenstein06} catalog with the RASS, we identified $22$ DAs listed in the \citet{mccook99} catalog with an X-ray source within $1'$. We classified seven of these sources as unlikely X-ray emitting WDs. In four cases, a known QSO is in the RASS error circle and the X-ray source is not super-soft, which is consistent with the QSOs being the X-ray sources. 

In three ambiguous cases, the sources are not super-soft and at large separations from the RASS source, but the available SDSS data do not rule out the DAs as X-ray sources. SDSS J022618.60$-$083049.3 is in a RASS error circle with two bright stars ($g = 12.98$ and $13.19$). The RASS source is not super-soft (HR1 $=0.22\pm0.36$) and it is therefore unlikely that the WD is the X-ray source. SDSS J093456.44$+$023121.1 is in a error circle with no obvious X-ray source counterparts. There is a very faint, red star closer to the RASS position ($g = 21.35, r = 20.06$), but were it the X-ray counterpart, its log $(f_X/f_g)$ would be much higher than typical for stars. The RASS source is not soft (HR1 $= 0.36\pm0.52$), so perhaps this is a QSO with stellar colors. Finally, SDSS J152349.89$+$041434.6 is in a error circle with SDSS J152349.88$+$041400.8, a $g = 17.71$ galaxy. In these cases follow-up spectroscopy is needed to reveal, for example, whether this galaxy hosts an AGN. We consider these three WDs to be less certain RASS counterparts and list them as such in Table~\ref{known_and_new_wds}.

Of the remaining $15$ previously known DAs, we include in Table~\ref{known_and_new_wds} the four that have not been identified as X-ray emitters (i.e., they are not in the \citet{fleming96}, \citet{zickgraf03}, or \citet{chu04} catalogs) as new identifications.\\

\noindent{\it SDSS DR4-discovered DAs}\\
\indent Our SDSS/RASS correlations yielded a further $25$ SDSS-discovered DA WDs (i.e., not listed in the \citet{mccook99} catalog) with an X-ray source within $1'$. Five of these sources also have nearby cataloged QSOs and are not super-soft. In three other cases we identify plausible non-WD counterparts to the RASS sources. A UV-excess source with $u-g = 0.00\pm0.03$ is positionally coincident with the RASS source for which SDSS J083300.12$+$272254.5 is proposed as a counterpart. Similarly, another UV-excess source (with $u-g = 0.24\pm0.13$) is closer than SDSS J094437.72$+$493815.3 to the RASS source for which the WD is the candidate counterpart. Finally, for SDSS J162814.59$+$363008.3 a very faint UV-excess object ($g = 21.45$, $u-g = 0.18\pm0.13$) is positionally coincident with a FIRST radio source. In all three cases the RASS sources are not super-soft, and it seems likely that the UV-excess objects are QSOs and X-ray sources. Follow-up spectroscopy is required to confirm this, but we consider these DAs to be unlikely RASS-source counterparts.

Of the $17$ remaining DAs, there are seven cases in which inspection of the RASS error circle, combined with the RASS source hardness ratio, suggests that the WD is probably not the X-ray source--but does not provide a strong alternate X-ray counterpart candidate. In all these error circles spectroscopy of multiple objects or high spatial resolution X-ray observations are needed. We consider these seven WDs to be less certain RASS counterparts and list them as such in Table~\ref{known_and_new_wds}.

The remaining $10$ SDSS DR4 DAs are listed in Table~\ref{known_and_new_wds} as new identifications.

\section*{Building the sample of X-ray emitting non-DAs}
Our SDSS/RASS correlations also returned $11$ non-DA WDs. Once again we eliminate those WDs in error circles with cataloged QSOs; there are three such error circles here, all with RASS sources that are not super-soft. For two other RASS sources the SDSS data suggest other counterparts. SDSS J123337.71+570335.4 is a z $= 0.351$ redshift QSO nearer to the RASS source (HR1 $=0.20\pm0.42$) than SDSS J123345.52$+$570343.9, a DB WD. The UV-excess object SDSS J155209.12$+$433821.8 ($u-g = 0.45\pm0.04$) is also closer to the RASS source (HR1 $=-0.18\pm0.37$) than SDSS J155205.59$+$433825.8, one of the DA+M-star binaries. While spectroscopic confirmation is needed in the latter case, we conclude that both of these RASS sources are QSOs.

The six remaining WDs are the likely RASS sources listed in Table~\ref{other_wds}. In addition to the two DBs discussed in \S~\ref{dbs}, these objects include the previously known PG 1159 star SDSS J114635.23$+$001233.5. The spectrum for this (very hot) star is poorly fit by our models, and the value for T$_{eff}$ in Table~\ref{other_wds} should therefore not be trusted. The five other stars are DA+M-star systems, of which one is an SDSS discovery, SDSS J223051.15$+$125706.8 \citep{silvestri06}. For the three systems included in the \citet{silvestri06} DR4 catalog of close binary systems, we give the published T$_{eff}$ and note the M stars' spectral types. For the two other systems, SDSS J132725.72$+$505711.0 (WD 1325$+$512) and J141740.22$+$130148.5 (WD 1415$+$132), we fit the spectra to obtain T$_{eff}$ (listed in Table~\ref{other_wds}) and log g ($8.93\pm0.09$ and $7.19\pm0.02$, respectively). The fit to the spectrum of SDSS J1327 is poor; the spectrum has relatively low signal-to-noise, and the presence of the M dwarf appears to have corrupted the fit to the H$\alpha$ line.

\section*{Building a sample of X-ray emitting WDs identified since DR4}
Since the production of the \citet{eisenstein06} WD catalog, new WDs have been identified through inspection of the more recent SDSS spectroscopy. We have also collected a few objects whose spectra were taken earlier in the survey but that were not included in the \citet{kleinman04} DR1 or \citet{eisenstein06} samples; we refer to all of these objects as post-DR4. 

\begin{figure*}
\epsscale{1.15}
\centerline{\plotone{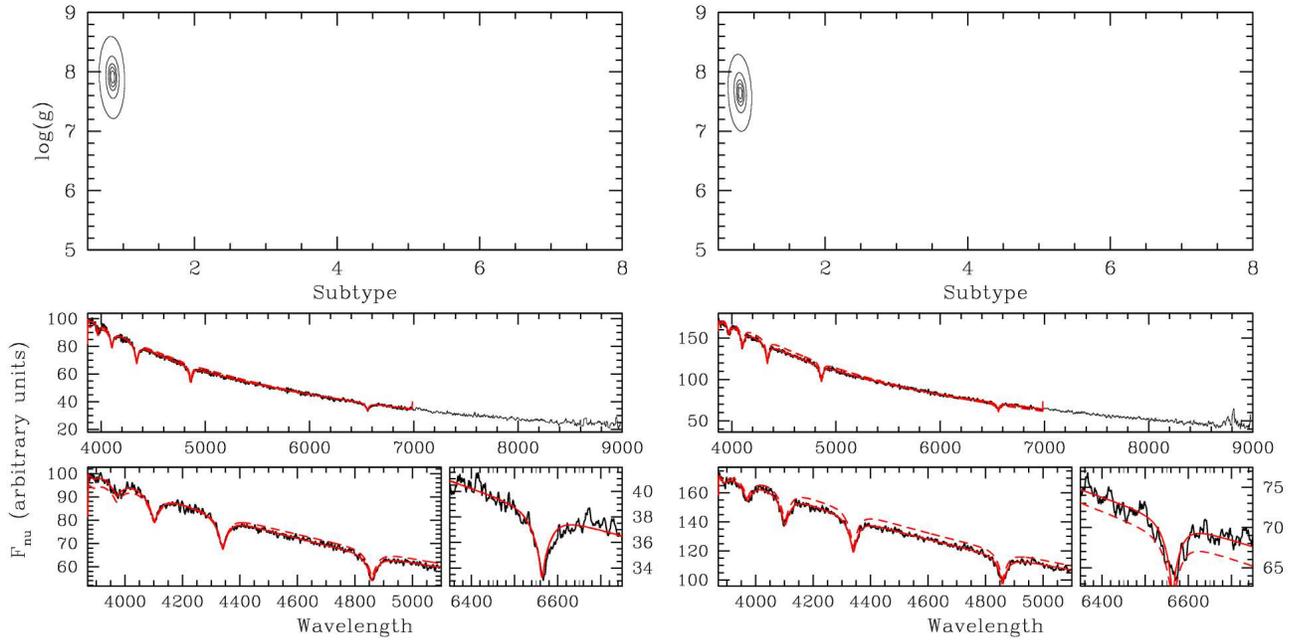}}
\caption{Output to the model fits to the spectra of SDSS J093921.83$+$264401.1 (left) and J100543.92$+$304744.9 (right). {\it Top panels:} fit contours as a function of the DA subtype, with subtype being T$_{eff}$ expressed in units of $50,400$ K/T$_{eff}$ \citep{sion83}. The likelihood contours are 1, 2, 3, 5, and 10 $\sigma$. {\it Bottom panels:} closeups of the fit plotted with the spectrum. The dashed line is the fit without reddening; the solid line is the fit with reddening taken into account.}\label{DA_fit1}
\end{figure*}

Among the $43$ RASS sources with new candidate WDs within $1'$, we selected the super-soft sources (HR1 $\leq -0.8$). Ten sources met this criterion. Two are known DA WDs, WD 1325$+$512 and WD 1415$+$132 \citep{mccook99}, and known X-ray emitters\footnote{WD 1415$+$132 is included in \citet{fleming96}; WD 1325$+$512 is listed in SIMBAD as a known X-ray source, but is not included in \citet{fleming96} or \cite {zickgraf03}. It is unclear how it was identified as a RASS source.}. The SDSS spectra indicate that both have M-star companions, and we include these WDs in Table~\ref{other_wds}.

In two other cases, the SDSS data indicate that another object may be the RASS counterpart. For SDSS J114142.84$+$143356.5, a $g = 15.76$ star is nearer to the position of the RASS  source (1RXS J14145.3$+$143349). This star, were it to be the X-ray source counterpart, would have log $(f_X/f_g) = -1.04$, which is high for a main sequence stellar X-ray emitter. On the other hand, this is the least soft of the RASS sources (HR1 $= -0.88\pm0.20$). For SDSS J155109.55$+$454252.2, a UV-excess object ($g = 17.55$, $u-g = -0.39\pm0.03$) is virtually coincident with the RASS source (1RXS J155108.6$+$454312). Since the RASS source is super-soft, this UV-excess object may be a WD or a CV; the latter are also frequently super-soft emitters.

In the remaining six cases the association of RASS source and SDSS candidate WD is very strong. Fitting the SDSS spectra using WD models then reveals that three of these candidate WDs are in fact hot stars: SDSS J125338.50$+$051542.1, J133024.24$-$001636.8, and J161614.02$+$080905.8. 

The other three stars are new DA WDs and included in Table~\ref{known_and_new_wds}: SDSS J093921.83$+$264401.1, J100543.92$+$304744.9, and J143736.67$+$362213.4. Because of calibration problems with the SDSS data, the fitting failed to return values for T$_{eff}$ and log g for the spectrum of SDSS J143736.67$+$362213.4 (the spectrum is also truncated at $7400$ \AA). The fits for the two other DAs are shown in Figure~\ref{DA_fit1}.

\end{document}